\documentclass[nofootinbib, prb,preprintnumbers,amsmath,amssymb,showkeys]{revtex4}
\usepackage{graphicx}
\usepackage{dcolumn}
\usepackage{bm}
\usepackage{comment}
\usepackage{float}
\usepackage{subfig}
\usepackage{epstopdf}

\begin{document}
%
\title{Excitonic phase transition in the extended three-dimensional\\Falicov-Kimball model}
%
\author{V. Apinyan\footnote{Tel.: +48 71 343 5021; fax: +48 71 344 1029.\newline \quad\quad 	E-mail address: V.Apinyan@int.pan.wroc.pl.}
}

\author{T. K. Kope\'{c}}
 
\affiliation{%
Institute for Low Temperature and Structure Research, Polish Academy of Sciences\\
PO. Box 1410, 50-950 Wroclaw 2, Poland \\}%

\date{\today}

\begin{abstract}
%
We study the excitonic phase transition in a system of the conduction band electrons and valence band holes described by the three-dimensional (3D) extended Falicov-Kimball (EFKM) model with the tunable Coulomb interaction $U$ between both species. By lowering the temperature, the electron-hole system may become unstable with respect to the formation of the excitons, i.e, electron-hole pairs at temperature $T=T_{\Delta}$, exhibiting a gap $\Delta$ in the particle excitation spectrum. To this end we implement the functional integral formulation of the EFKM, where the Coulomb interaction term is expressed in terms of U(1)  phase variables conjugate to the local particle number, providing a useful representation of strongly correlated system. The effective action formalism allows us to formulate a problem in the phase-only action in the form of the quantum rotor model and to obtain analytical formula for the critical lines and other quantities of physical interest like charge gap, chemical potential and the correlation length.
\end{abstract}
\maketitle

\section{\label{Section_1} Introduction}
%
The Coulomb interaction between the conduction band electrons and the valence band holes causes in some solid state materials the formation of the new bound states called the excitons.\cite{Moskalenko} The low-density system of excitons behaves like a weakly non-ideal Bose-gas.\cite{Keldysh_1} These new formations lead to the various interesting physical phenomena in solid state materials and they are the subjects of the intensive experimental and theoretical researches.\cite{Neuenschwander,Wachter_1,Wachter_2,Wachter,Keldysh_2,Cloizeaux,Kohn,Jerome, Keldysh_3,Snoke_1,Snoke_2} In the scenario of the semiconductor-metal phase transition, a new phase develops approaching to the transition from the semiconductor side.\cite{Jerome} This state is called as the ``excitonic insulator''\cite{Halperin} (EI) and is characterized by the strong binding between the conduction band electrons and valence band holes. For example, the series of recent experimental investigations \cite{Neuenschwander, Wachter, Wachter_1,Wachter_2} in TmSe$_{0.45}$Te$_{0.55}$ have suggested the existence of the EI state in that material. Another example of the material with a well defined EI state is the quasi-one-dimensional Ta$_{2}$NiSe$_{5}$ with highly polarizable Se. The angle-resolved photoemission spectra (ARPES) on these compounds \cite{Wakisaka} demonstrate that the ground state therein is an excitonic insulator. The evidence in favor of the EI state is proved also in the transition metal layered compound $1T$-TiSe$_{2}$, \cite{Berger} where the EI scenario is driving to the charge-density wave transition in such a material.

Turning to the theory, there have been a number of works on the excitonic systems. Using the band structure calculation and the mean-field (MF) analysis for the EI state \cite{Kaneko} it was found that a structural phase transition driven by the exciton Bose-Einstein condensation (BEC) takes place in the layered chalcogenide material such as the recently corroborated sample of Ta$_{2}$NiSe${_5}$. In the small interaction region, the system is in the Bardeen-Cooper-Schrieffer (BCS) state\cite{Bardeen} with weakly bound electron-hole pairs, while, approaching from the semiconductor side, the system shows typical BEC behavior with tightly bound excitons,\cite{Ihle,Pethick} thus exhibiting a BCS-BEC type crossover.\cite{Chen} This type of crossover mechanism is found in a study of the electron-hole plasma condensation in highly excited semiconductors.\cite{Kremp} In another work, a BEC-BCS crossover was studied using the effective-mass model, for the valence band holes and conduction band electrons.\cite{Bronold} In this context, the three-dimensional (3D) extended spinless Falicov-Kimball model (EFKM) with the dispersive $f$-orbital electrons at half-filling has been analyzed recently.\cite{Zenker_1, Zenker_2} The spontaneous symmetry breaking for the EI state and BCS-BEC like crossover for the 2D extended Falicov-Kimball model is discussed also in Ref.\ \onlinecite{Seki}. The spectrum of low-energy collective excitations in the EFKM is discussed recently.\cite{Golosov} The MF stability of the EI state observed within the EFKM model is attributed to the broken degeneracy, due the presence of the finite $f$-band hopping. It is shown that the EI state is unstable when the case of the pure Falicov-Kimball model (FKM) (fully localized bands) is
approached. Also the Bogoliubov-de Gennes equations were implemented using the exact diagonalization method, and the Hartree-Fock (HF) type self-consistent equations for the ground state of the spinless EFKM model are derived in two and three dimensions. \cite{Farkasovsky_1} Based on the analysis of electron-hole pairing in the extended Falicov-Kimball model, the authors in Ref.\ \onlinecite{Zenker_3} show that tuning the Coulomb attraction between both species, a continuous BCS-BEC like crossover might be achieved. Moreover, it has been shown that the $f$-$f$ hopping mechanism could be also responsible for the exciton formation.\cite{Batista_1, Batista_2, Czycholl}

The importance of the phase coherence in the excitonic pair (EP) plasma is discussed recently, \cite{Snoke_1, Snoke_2} where a classification of two distinct transitions in the excitonic plasma is given and discussion about the exciton condensation conditions is provided. Particularly, it is shown theoretically that the excitonic insulator and the excitonic condensate are not exactly the same.\cite{Snoke_1, Snoke_2, Tomio} The author in Refs.\ \onlinecite{Snoke_1} and \onlinecite{Snoke_2} shows from general considerations that in the low density limit of the excitonic pairs, the critical temperature $T_{c}$ of excitonic BEC should be much smaller than the temperature $T_{\Delta}$ of the EP formation. This is in contrast with tprevious treatments,\cite{Ihle, Zenker_1, Zenker_2, Zenker_3} where the 
EI state is associated with the BEC state of excitons. Similarly, in Ref.\ \onlinecite{Tomio} it is shown that the EI state is an excitonium state, where the incoherent e-h bound pairs are formed and furthermore, at the lower temperatures, the BEC of excitons appears in consequence of the reconfiguration and coherent condensation of preformed excitonic pairs. Obviously, in the low density limit, the gas of free excitons undergoes the BEC phase transition at the very low temperatures, and the BEC temperature transition line is not coinciding with that of EP formation. The Bose condensation of the excitonic pairs is possible only when the macroscopic phase coherence is present in the system.\cite{Snoke_1} 

Contrary, at high e-h density, where the mean distance between the particles is shorter than the excitonic Bohr radius, the weakly bound e-h pairs behave like the Cooper pairs in the conventional superconductors and at sufficiently low temperatures, i.e., the BCS state of e-h pairs.\cite{Keldysh_2, Jerome, Micnas} Therefore, an expected BCS-BEC crossover, represents actually a fascinating problem typical to the excitonic systems. Especially, it is interesting from the viewpoint of the difference from similar crossover in superconductors, or the trapped atomic Fermi gases. \cite{Micnas, Randeria, Ohashi} The transition to e-h pair condensed phase, in the weak-coupling limit, is related to the relative motion between electrons and holes,\cite{Tomio} implying the BCS regime and is in contrast to the case of strong-coupling regime, when the BEC state is related to the motion of the center of mass of excitons. In the whole BCS - BEC transition region, the e-h mass difference
leads to a large suppression of the BEC transition temperature, which is proved to not be same as EP formation temperature. \cite{Tomio} 

In the present paper we explore the quantum collective  behavior of the excitons
in 3D system going beyond the simple HF method.
To this end, we study the excitonic phase transition in a system composed of the conduction band electrons and valence band holes, described by the 3D extended Falicov-Kimball \cite{Falicov, Ramirez_1} model with tunable Coulomb interaction $U$ between both species of particles. We implement the quantum rotor approach, where the Coulomb interaction of the EFKM model  is expressed in terms of U(1) quantum phase variables conjugate to the local particle number, providing a useful representation of strongly correlated systems. 
This allows us to obtain the analytical formulas for the critical lines and other quantities of physical interest like the charge gap, chemical potential and the correlation length. We present also the numerical evaluations of all physical quantities discussed in the paper.

The plan of the paper is as follows: in the Section \ref{sec:Section_2} we provide the Hamiltonian of the model EFKM, then in the Section \ref{sec:Section_3} we introduce the new decoupling potentials and we handle with four fermion interaction term in the initial Hamiltonian. Furthermore, in the Section \ref{sec:Section_4}, we obtain the transition temperature of the excitonic pair formation, excitonic gap parameter, the charge gap and other important physical quantities. At the end of that section we discuss the numerical results. In Section \ref{sec:Section_5} we obtain the effective phase action in the context of the quantum rotor approach and we derive the equation for the excitonic BEC transition critical temperature. Numerical results are also discussed there. The momentum distribution functions and the excitonic coherence length are calculated in the \ref{sec:Section_6}. Conclusions are given in Section \ref{sec:Section_6}. A number of technical details is given in Appendices.
\newline
%
\section{\label{sec:Section_2} The Hamiltonian}
%
We consider the Hamiltonian of the extended Falicov-Kimball model
\begin{eqnarray}
&&{\cal{H}}=-t\sum_{\left\langle {\bf{r}},{\bf{r}}' \right\rangle}\left[\bar{c}({{\bf{r}}})c({{\bf{r}}}')+h.c.\right]-\left(\mu-\epsilon_{c}\right)\sum_{{\bf{r}}}n_{c}({\bf{r}})
\nonumber\\
&&-\tilde{t}\sum_{\left\langle {\bf{r}},{\bf{r}}' \right\rangle}\left[\bar{f}({{\bf{r}}})f({{\bf{r}}}')+h.c.\right]-\left(\mu-\epsilon_{f}\right)\sum_{{\bf{r}}}n_{f}({\bf{r}})
\nonumber\\
&&+U\sum_{{\bf{r}}}n_{c}({{\bf{r}}})n_{f}({{\bf{r}}}),
\label{Equation_1}
\end{eqnarray}
where $\bar{c}({{\bf{r}}})$ (${c}({{\bf{r}}})$) are the creation (annihilation) operators of the electron of the $c$-orbitals at the site with the position ${\bf{r}}$ and $\left\langle {\bf{r}} {\bf{r}}' \right\rangle$ runs over pairs of the nearest neighbor (n.n.) sites on a 3D cubic lattice. Furthermore $t$ is the hopping integral for the $c$-electrons and $\epsilon_{c}$ is the on-site energy level. Similarly, $\bar{f}({{\bf{r}}})$ (${f}({{\bf{r}}})$) are the creation (annihilation) operators of the  $f$-orbital electrons and $\tilde{t}$ is the hopping integral for the $f$-electrons. The EFKM Hamiltonian in Eq.(\ref{Equation_1}) is equivalent to the asymmetric Hubbard model, if we associate to the orbitals $c$ and $f$ the spin variables, thus replacing the fermion Hilbert-space by the pseudo-fermionic one, and by linearizing the interaction term via bosonic states. \cite{Zenker_1} Furthermore, $\epsilon_{f}$ is the on-site energy level of the $f$-orbital and  $\mu$ is the chemical potential. The equilibrium value of chemical potential $\mu$ will be determined from the half-filling condition, i.e., $n_{c}=1- n_{f}$, where $n_{x}\equiv\left\langle n_{x}({\bf{r}})\right\rangle$ is the average particle density with $x=c, f$ for the $c$ and $f$-orbital electrons respectively. Furthermore, we suppose that the chemical potentials of both orbitals are the same, as in the work in Ref.\ \onlinecite{Seki}. The parameter $U$, which enters in the last term of the Hamiltonian, is the Coulomb repulsion between two types of electrons. Furthermore $n_{c}({\bf{r}})$ and $n_{f}({\bf{r}})$ are the $c$- and $f$-electron density operators and they are defined as usual by the relation $n_{x}({\bf{r}})=\bar{x}({\bf{r}})x({\bf{r}})$.

We consider also the following values for the band parameters $\epsilon_{c}=0$ and $\epsilon_{f}=-1$. With this consideration the $c$- electrons are itinerant and the $f$-electrons are quasilocalized on the atomic sites. Throughout the paper we set $k_{B} = 1$ and $\hbar=1$ and lattice constant $a=1$. 
%
\section{\label{sec:Section_3} The method}
%
In the first step, we transform the fermionic interaction term in the Hamiltonian by rewriting the density product in the last term in Eq.(\ref{Equation_1}) in the equivalent form
\begin{eqnarray}
n_{c}({{\bf{r}}})n_{f}({{\bf{r}}})=\frac{n^{2}({\bf{r}})}{4}-\frac{\tilde{n}^{2}({\bf{r}})}{4},
\label{Equation_2}
\end{eqnarray}
where we introduced the short-hand notations 
\begin{eqnarray}
n({\bf{r}})=n_{c}({\bf{r}})+n_{f}({\bf{r}}),
\newline\\\tilde{n}({\bf{r}})=n_{c}({\bf{r}})-n_{f}({\bf{r}}).
\label{Equation_3}
\end{eqnarray}
With the new notations we can rewrite the Hamiltonian of the system in Eq.(\ref{Equation_1})  as
\begin{eqnarray}
&&{\cal{H}}=-t\sum_{\left\langle {\bf{r}},{\bf{r}}' \right\rangle}\left[\bar{c}({{\bf{r}}})c({{\bf{r}}}')+h.c.\right]-\bar{\mu}\sum_{{\bf{r}}}n({\bf{r}})
\nonumber\\
&&-\tilde{t}\sum_{\left\langle {\bf{r}},{\bf{r}}' \right\rangle}\left[\bar{f}({{\bf{r}}})f({{\bf{r}}}')+h.c.\right]+\frac{\epsilon_{c}-\epsilon_{f}}{2}\sum_{{\bf{r}}}\tilde{n}({\bf{r}})
\nonumber\\
&&+U\sum_{{\bf{r}}}\frac{1}{4}\left[n^{2}({\bf{r}})-\tilde{n}^{2}({\bf{r}})\right].
\label{Equation_4}
\end{eqnarray}
We have putted here $\bar{\mu}=\mu-\bar{\epsilon}$ and $\bar{\epsilon}=\left(\epsilon_{c}+\epsilon_{f}\right)/2$ is the average  energy level parameter. The Hamiltonian in Eq.(\ref{Equation_4}) is now suitable for decoupling quadratic density terms using the Gaussian path integral method. \cite{Negele}
%
\subsection{\label{sec:Section_3_1} Functional integral formalism: decoupling of interactions}
%
Dealing with fermions within the path integral method, requires introduction of the Grassmann variables ${c}({{\bf{r}}}\tau)$ and ${f}({{\bf{r}}}\tau)$ at each site ${\bf{r}}$ and at each imaginary time $\tau$. The latest is varying in the interval $0\leq \tau \leq\beta$, where $\beta=1/T$ (with $T$ being the thermodynamic temperature). The variables ${c}({{\bf{r}}}\tau)$ and ${f}({{\bf{r}}}\tau)$ satisfy the anti-periodic boundary conditions ${x}({{\bf{r}}}\tau)=-{x}({{\bf{r}}}\tau+\beta)$. 
The partition function of system of the fermions, written as a functional integral over the Grassmann field, is
\begin{eqnarray}
{\cal{Z}}=\int \left[{\cal{D}}\bar{c}{\cal{D}}c\right]\int\left[{\cal{D}}\bar{f}{\cal{D}}f\right]e^{-{\cal{S}}[\bar{c},c, \bar{f},f]},
\label{Equation_5}
\end{eqnarray} 
where the action in exponential is given in the path integral formulation as
\begin{eqnarray}
{\cal{S}}[\bar{c},c, \bar{f},f]=\sum_{x=c,f}{\cal{S}}_{B}[\bar{x},x]+\int^{\beta}_{0}d\tau {\cal{H}}(\tau).
\label{Equation_6}
\end{eqnarray} 
Here ${\cal{S}}_{B}[\bar{x},x]$ is the fermionic Berry term for the $c$ and $f$-electrons. It is defined as
\begin{eqnarray}
{\cal{S}}_{B}[\bar{x},x]=\sum_{{\bf{r}}}\int^{\beta}_{0}d\tau \bar{x}({\bf{r}}\tau)\frac{\partial}{\partial{\tau}}x({\bf{r}}\tau).
\label{Equation_7}
\end{eqnarray}
Next, we decouple quadratic density terms in Eq.(\ref{Equation_4}) using the Hubbard-Stratonovich (HS) transformation \cite{Negele} and by introducing the new variables $V({\bf{r}}\tau)$ and ${\cal{ \varrho}}({\bf{r}}\tau)$ conjugated to the density terms $n({\bf{r}}\tau)$ and $\tilde{n}({\bf{r}}\tau)$ respectively. 
For the quadratic term proportional to $n^{2}({\bf{r}}\tau)$, in the exponential of the partition function in Eq.(\ref{Equation_5}), we have 
\begin{eqnarray}
&&\exp\left[{-\frac{U}{4}\sum_{{\bf{r}}}\int^{\beta}_{0}d\tau n^{2}\left({\bf{r}}\tau\right)}\right]
\nonumber\\
&&=\int\left[{\cal{D}}V\right]e^{-\sum_{{\bf{r}}}\int^{\beta}_{0}d\tau \left[\frac{V^{2}({\bf{r}}\tau)}{U}-iV({\bf{r}}\tau)n({\bf{r}}\tau)\right]}.
\nonumber\\
\ \ \ 
\label{Equation_8}
\end{eqnarray} 

After combining the exponential in Eq.(\ref{Equation_8}) with the term linear in total electron density $n({\bf{r}})$ in Eq.(\ref{Equation_4}), we can decompose the variables $V({\bf{r}}\tau)$ into the static and periodic parts
\begin{eqnarray}
V({\bf{r}}\tau)=V_{0}({\bf{r}})+\tilde{V}({\bf{r}}\tau),
\label{Equation_9}
\end{eqnarray}
where $\int^{\beta}_{0}d\tau \tilde{V}({\bf{r}}\tau)=0$. As a result, the integration over $V({\bf{r}}\tau)$-variables becomes now the integration over the scalar static variables $V_{0}({\bf{r}})$ and the integration over the periodic field $\tilde{V}({\bf{r}}\tau)$:
\begin{eqnarray}
\int\left[{\cal{D}}V\right]...=\int\left[{\cal{D}}V_{0}\right]\int\left[{\cal{D}}\tilde{V}\right]... \; .
\label{Equation_10}
\end{eqnarray}
For the periodic part in Eq.(\ref{Equation_9}), we introduce the U$(1)$ phase field variables $\phi({\bf{r}}\tau)$ using a Faraday-type relation \cite{Kopec_1}
\begin{eqnarray}
\tilde{V}({\bf{r}}\tau)=\frac{\partial{\varphi}({{\bf{r}}}\tau)}{\partial{\tau}}\equiv \dot{\phi}({\bf{r}}\tau).
\label{Equation_11}
\end{eqnarray}
Thus, for the dynamic part, we transform the integration over the gauge variables $\tilde{{V}}({\bf{r}}\tau)$ into the integration over the generic phase variables $\varphi({\bf{r}}\tau)$
\begin{eqnarray}
\int\left[{\cal{D}}\tilde{V}\right]... \rightarrow \int\left[{\cal{D}}\phi\right]... .
\label{Equation_12}
\end{eqnarray}
The periodicity of $\tilde{V}\left({\bf{r}}\tau\right)$ implies that $\phi\left({\bf{r}}\beta\right)=\phi\left({\bf{r}}0\right)$.
The integration measure in Eq.(\ref{Equation_12}) over $\phi$ variables is defined as
\begin{eqnarray}
\int\left[{\cal{D}}\phi\right]...\equiv \int^{\infty}_{-\infty}\prod_{{\bf{r}}}d\phi_{0}({\bf{r}})
\nonumber\\
\times \int^{\phi_{f}=\phi({\bf{r}}\beta)}_{\phi_{i}=\phi_{0}({\bf{r}})}\prod_{{\bf{r}}}d\phi({\bf{r}}\tau)... , 
\label{Equation_13}
\end{eqnarray}
where the notations $\phi_{i}$ and $\phi_{f}$ mean the initial and final paths.
The path integral in Eq.(\ref{Equation_13}) could be transformed into path integration over the compact U(1) group manifold, since the electromagnetic U(1) group governing the phase field
is compact, i.e. $\phi({\bf{r}}\tau)$ has the topology of a circle ($S_1$), thus we have a non-homotopic mapping of the configuration space onto the U(1) gauge group $S_{1}\rightarrow U(1)$. The paths, which
loop around a circle in different number of times, are in different
homotopy classes and they cannot be continuously deformed
into one another. All these paths can be characterized by their proper winding numbers $m\left({\bf{r}}\right)$. Any two paths, which have different winding numbers, cannot be continuously  transformed one to another, and in order to include all possible phase path contributions, we have to sum
over all topologically inequivalent phase configurations described
by their winding numbers. Accordingly, the path integral in Eq.(\ref{Equation_13}) is transformed as
\begin{eqnarray}
\int\left[{\cal{D}}\phi\right]...=\int\left[{\cal{D}}\varphi\right]... \; .
\label{Equation_14}
\end{eqnarray}
The integration measure in Eq.(\ref{Equation_14}) is now
\begin{eqnarray}
\int\left[{\cal{D}}\varphi\right]...\equiv \sum_{\left\{m({\bf{r}})\right\}}\int^{2\pi}_{0}\prod_{{\bf{r}}}d\varphi_{0}({\bf{r}})
\nonumber\\
\times \int^{\varphi({\bf{r}}\beta)=\varphi_{0}\left({\bf{r}}\right)+2{\pi}m({\bf{r}})}_{\varphi\left({\bf{r}}0\right)=\varphi_{0}\left({\bf{r}}\right)}\prod_{{\bf{r}}}d\varphi({\bf{r}}\tau)... \; 
\label{Equation_15}
\end{eqnarray}
In performing the integration over the phase field one should take into account that the field configurations satisfy the boundary conditions \cite{Kopec_2,Trombettoni} 
\begin{eqnarray}
\varphi({\bf{r}}\beta)-\varphi({\bf{r}}0)=2\pi{m({\bf{r}})}.
\label{Equation_16}
\end{eqnarray}
Thus, integration over all phases $\phi({\bf{r}}\tau)$ amounts the integration over the $\beta$-periodic field $\varphi({\bf{r}}\tau)$ and the summation over a set of U(1) winding numbers $m({\bf{r}})$.
For the scalar static part $V_{0}({\bf{r}})$, we have the following functional integral
\begin{eqnarray}
&&\int\left[{{d}}V_{0}\right]e^{\sum_{{\bf{r}}}\int^{\beta}_{0}d\tau -\frac{V^{2}_{0}({\bf{r}})}{U}+iV_{0}({\bf{r}})\left[n({\bf{r}}\tau)-\frac{2\bar{\mu}}{U}\right]}.
\label{Equation_17}
\end{eqnarray}
The saddle-point value of ${V}^{s.p.}_{0}({{\bf{r}}})$ is given as ${{V}^{s.p.}_{0}}=i\frac{Un}{2}-i\bar{\mu}$,
where $n$ is total average particle density $n=n_{c}+n_{f}$. 
And we have the contribution in the partition function in Eq.(\ref{Equation_5}) in the form
\begin{eqnarray}
\exp\left[{-{\cal{S}}\left[\varphi\right]-\sum_{{\bf{r}}}\int^{\beta}_{0}d\tau{\mu}_{n}n({\bf{r}}\tau)}\right],
\label{Equation_18}
\end{eqnarray}
where the effective phase-only action ${\cal{S}}[\varphi]$ is given as 
\begin{eqnarray}
{\cal{S}}[\varphi]=\sum_{{\bf{r}}}\int^{\beta}_{0}d\tau\left[\frac{\dot{\varphi}^{2}({\bf{r}}\tau)}{U}-\frac{2\bar{\mu}}{iU}\dot{\varphi}({\bf{r}}\tau)-i\dot{\varphi}({\bf{r}}\tau)n({\bf{r}}\tau)\right]
\nonumber\\
\ \ \ 
\label{Equation_19}
\end{eqnarray}
and the effective chemical potential ${\mu}_{n}$ attached to the total density operator is introduced as ${\mu}_{n}=\frac{Un}{2}-\bar{\mu}$.

The decoupling of the quadratic term proportional to $\tilde{n}^{2}({\bf{r}}\tau)$ in the exponential of the partition function in Eq.(\ref{Equation_5}) is also straightforward. We obtain
\begin{eqnarray}
&&\exp\left[{\sum_{{\bf{r}}}\int^{\beta}_{0}d\tau \frac{U}{4}\tilde{n}^{2}({\bf{r}}\tau)}\right]
\nonumber\\
&&=\int\left[{\cal{D}}{\varrho}\right]e^{-\sum_{{\bf{r}}}\int^{\beta}_{0}d\tau \left[\frac{\varrho^{2}({\bf{r}}\tau)}{U}-\varrho({\bf{r}}\tau)\tilde{n}({\bf{r}}\tau)\right]}.
\nonumber\\
\ \ \ 
\label{Equation_20}
\end{eqnarray}
By combining the expression in the exponential in Eq.(\ref{Equation_20}) with the similar term, linear in $\tilde{n}$, in the expression of the transformed Hamiltonian in Eq.(\ref{Equation_4}) and then by shifting the integration variables, we get
\begin{eqnarray}
&&\int\left[{\cal{D}}{\varrho}\right]e^{\sum_{{\bf{r}}}\int^{\beta}_{0}d\tau-\frac{\varrho^{2}({\bf{r}}\tau)}{ U}+\varrho({\bf{r}}\tau)\left[\tilde{n}({\bf{r}}\tau)-\frac{\epsilon_{c}-\epsilon_{f}}{2U}\right]}.
\nonumber\\
\label{Equation_21}
\end{eqnarray}
The saddle-point evaluation gives for $\varrho$
\begin{eqnarray}
\varrho^{s.p.}_{0}=\frac{U\tilde{n}}{2}-\frac{\epsilon_{c}-\epsilon_{f}}{2},
\label{Equation_22}
\end{eqnarray}
where $\tilde{n}=\left\langle \tilde{n}({\bf{r}}\tau)\right\rangle$.
As the result of decoupling, we obtain following ``Zeeman-like'' contribution in the partition function
\begin{eqnarray}
\exp\left[{-\sum_{{\bf{r}}}\int^{\beta}_{0}d\tau\mu_{\tilde{n}}\tilde{n}({\bf{r}}\tau)}\right]
\label{Equation_23}
\end{eqnarray}
with attached effective chemical potential $\mu_{\tilde{n}}=\frac{\epsilon_{c}-\epsilon_{f}}{2}-\frac{U\tilde{n}}{2}$.

To summarize, the partition function of the system after the decoupling procedures will be
\begin{eqnarray}
{\cal{Z}}=\int \left[{\cal{D}}\bar{c}{\cal{D}}c\right]\left[{\cal{D}}\bar{f}{\cal{D}}f\right]\left[{\cal{D}}\varphi\right]e^{-{\cal{S}}[\bar{c},c,{\bar{f}},f,\varphi]},
\label{Equation_24}
\end{eqnarray}
where the action ${\cal{S}}[\bar{c},c,{\bar{f}},f,\varphi]$ in the exponential is 
\begin{eqnarray}
&&{\cal{S}}[\bar{c},c,{\bar{f}},f,\varphi]={\cal{S}}\left[\varphi\right]+\sum_{x=c,f}{\cal{S}}_{B}[\bar{x},x]
\nonumber\\
&&-t\sum_{\left\langle{\bf{r}},{\bf{r}}' \right\rangle}\int^{\beta}_{0}d\tau \left[\bar{c}({{\bf{r}}}\tau)c({{\bf{r}}}'\tau)+h.c.\right]
\nonumber\\
&&-\tilde{t}\sum_{\left\langle {\bf{r}},{\bf{r}}' \right\rangle}\int^{\beta}_{0}d\tau \left[\bar{f}({{\bf{r}}}\tau)f({{\bf{r}}}'\tau)+h.c.\right]
\nonumber\\
&&\ +\sum_{{\bf{r}}}\int^{\beta}_{0}d\tau \left[{\mu}_{n}n({\bf{r}}\tau)+\mu_{\tilde{n}}\tilde{n}({\bf{r}}\tau)\right].
\label{Equation_25}
\nonumber\\
\end{eqnarray}
The action in the form given in Eq.(\ref{Equation_25}) is suitable for derivation of the effective phase action and the fermionic action.
%
\subsection{\label{sec:Section_3_2} The U(1) gauge transformation}
%
In the perspective of treating the local and non-local correlations in the excitonic system it is important to separate the U(1) gauge degrees of freedom related to the phase sector. To this end, we perform the local gauge transformation to new fermionic Grassmann variables $a({\bf{r}}\tau)$ and $b({\bf{r}}\tau)$. This procedure will automatically eliminates also the last imaginary term appearing in the expression of the phase action in Eq.(\ref{Equation_19}). For the electrons of $c$ and $f$-orbitals the U$(1)$ transformation is 
\begin{eqnarray}\left[
\begin{array}{cc}
x({\bf{r}}\tau) \\
\bar{x}({\bf{r}}\tau)
\end{array}
\right]=\hat{{\cal{U}}}(\varphi)\cdot\left[
\begin{array}{cc}
\tilde{x}({\bf{r}}\tau) \\
\bar{\tilde{x}}({\bf{r}}\tau)
\end{array}
\right],
\label{Equation_26}
\end{eqnarray}
where $\hat{\cal{U}}(\varphi)$ is the U(1) transformation matrix $\hat{\cal{U}}(\varphi)=\hat{I}\cdot\cos\varphi({\bf{r}}\tau)+i\hat{\sigma}_{z}\cdot\sin\varphi({\bf{r}}\tau)$ with the unit matrix $\hat{I}$ and $\hat{\sigma}_{z}$ being the Pauli matrix. The variables $\tilde{x}=a$, $b$.  
We used the bosonic phase variables $\varphi\left({\bf{r}}\tau\right)$ introduced in Eq.(\ref{Equation_14}). In fact, the electron factorization in terms of two variables has an unprecedented impact on the whole theory. Especially, the emergent bosonic gauge sector, related to the phase variables, leads to a Bose-type of band bandwidth-renormalization factor (see in the Section \ref{sec:Section_4}).

The action in the Eq.(\ref{Equation_25}) after transformation procedure takes the following form 
\begin{eqnarray}
&&{\cal{S}}[\bar{a},a,{\bar{b}}, b,\varphi]={\cal{S}}_{0}[\varphi]+\sum_{\tilde{x}=a,b}{\cal{S}}_{B}[\bar{\tilde{x}},\tilde{x}]
\nonumber\\
&&-t\sum_{\left\langle{\bf{r}},{\bf{r}}' \right\rangle}\int^{\beta}_{0}d\tau \left[\bar{a}({{\bf{r}}}\tau)a({{\bf{r}}}'\tau)e^{-i\left[\varphi({{\bf{r}}}\tau)-\varphi({{\bf{r}}}'\tau)\right]}+h.c.\right]
\nonumber\\
&&-\tilde{t}\sum_{\left\langle {\bf{r}},{\bf{r}}' \right\rangle}\int^{\beta}_{0}d\tau \left[\bar{b}({{\bf{r}}}\tau)b({{\bf{r}}}'\tau)e^{-i\left[\varphi({{\bf{r}}}\tau)-\varphi({{\bf{r}}}'\tau)\right]}+h.c.\right]
\nonumber\\
&&\ \ \ +\sum_{{\bf{r}}}\int^{\beta}_{0}d\tau \left[{\mu}_{n}n({\bf{r}}\tau)+\mu_{\tilde{n}}\tilde{n}({\bf{r}}\tau)\right]
\nonumber\\
&&\ \ \ 
\label{Equation_27}
\end{eqnarray}
with the new phase action ${\cal{S}}_{0}[\varphi]$
\begin{eqnarray} {\cal{S}}_{0}[\varphi]=\sum_{{\bf{r}}}\int^{\beta}_{0}d\tau\left[\frac{\dot{\varphi}^{2}({\bf{r}}\tau)}{U}-\frac{2\bar{\mu}}{iU}\dot{\varphi}({\bf{r}}\tau)\right].
\label{Equation_28}
\end{eqnarray}
Then the partition function of the system in new variables is
\begin{eqnarray}
{\cal{Z}}=\int\left[{\cal{D}}\bar{a}{\cal{D}}a\right]\left[{\cal{D}}\bar{b}{\cal{D}}b\right]\left[{\cal{D}}\varphi\right] e^{-{\cal{S}}[\bar{a},a,{\bar{b}}, b,\varphi]}.
\label{Equation_29}
\end{eqnarray}
This form of the partition function will be the starting point for deriving the effective actions for the fermions and for the phase sector. 
%
%
\section{\label{sec:Section_4} Excitonic gap }
%
The EI low-temperature phase is characterized by local excitonic order parameter (excitonic gap).
The nonvanishing of the expectation value
\begin{eqnarray}
\Delta=U\left\langle \bar{a}({{\bf{r}}}\tau)b({{\bf{r}}}\tau)\right\rangle
\label{Equation_30}
\end{eqnarray}
signals the appearance of the electron-hole bound pairs, which manifests as a gap in the excitation spectrum and signals the presence of the EI state. The EI state develops from the local on-site electron-hole correlations.

We start with derivation of the EI gap equation. First, we apply the tranformation given in Eq.(\ref{Equation_26}) to fermionic variables in the initial Hamiltonian of the system in Eq.(\ref{Equation_1}). Then, we decouple four fermionic interaction term within the HF approach \cite{Czycholl} by applying Bogoliubov MF approximation. We have
\begin{eqnarray}
&&n_{a}({{\bf{r}}}\tau)n_{b}({{\bf{r}}}\tau)\approx\left\langle n_{a}({{\bf{r}}}\tau)\right\rangle n_{b}({{\bf{r}}}\tau)+\left\langle n_{b}({{\bf{r}}}\tau)\right\rangle n_{a}({{\bf{r}}}\tau)
\nonumber\\
&&-\frac{1}{U}\bar{\Delta}\bar{a}({{\bf{r}}}\tau)b({{\bf{r}}}\tau)-\frac{1}{U}{\Delta}\bar{b}({{\bf{r}}}\tau)a({{\bf{r}}}\tau).
\label{Equation_31}
\end{eqnarray}
Here $n_{a}({\bf{r}}\tau)$ and $n_{b}({\bf{r}}\tau)$ are the electron densities after the U$(1)$ gauge transformation.

The Fourier transformation of fermionic variables $a({\bf{r}}\tau)$ and $b({\bf{r}}\tau)$ is given by 
\begin{eqnarray}
x({\bf{r}}\tau)=\frac{1}{{\beta{N}}}\sum_{{\bf{k}},\nu_{n}}x_{{\bf{k}}}(\nu_{n})e^{i({\bf{k}}{\bf{r}}-\nu_{n}\tau)}
\label{Equation_32}
\end{eqnarray}
with $x=a,b$ for the $a$ and $b$ type electrons.
$N$ is the number of lattice sites and $\nu_{n}={\pi(2n+1)}/{\beta}$ are the Fermi-Matsubara frequencies with $n=0,\pm1,\pm2,...$. Furthermore, we will integrate out the phase variables in the expression of the partition function given in Eq.(\ref{Equation_29}), we obtain
\begin{eqnarray}
{\cal{Z}}=\int \left[{\cal{D}}\bar{a}{\cal{D}}a\right]\left[{\cal{D}}\bar{b}{\cal{D}}b\right]e^{-{\cal{S}}_{\rm eff}[\bar{a},a,{\bar{b}},b]},
\label{Equation_33}
\end{eqnarray}
where the effective phase-averaged fermionic action in the exponential is given by 
\begin{eqnarray}
{\cal{S}}_{\rm eff}[\bar{a},a,{\bar{b}},b]=-\ln{\int\left[{\cal{D}}\varphi\right]}e^{-{\cal{S}}[\bar{a},a,{\bar{b}},b,\varphi]}.
\label{Equation_34}
\end{eqnarray}
Now, using Eq.(\ref{Equation_32}) we can write the action ${\cal{S}}_{\rm eff}[\bar{a},a,{\bar{b}},b]$ in the Fourier space
\begin{eqnarray}
{\cal{S}}_{\rm eff}\left[{\bar{a},a,\bar{b},b}\right]&&=\frac{1}{\beta{N}}\sum_{{\bf{k}},\nu_{n}}\bar{a}_{{\bf{k}}}(\nu_{n})\left({\mu}^{a}_{\rm eff}-i\nu_{n}-{t}_{{\bf{k}}}\right)a_{{\bf{k}}}(\nu_{n})
\nonumber\\
&&+\frac{1}{\beta{N}}\sum_{{\bf{k}},\nu_{n}}\bar{b}_{{\bf{k}}}(\nu_{n})\left({\mu}^{b}_{\rm eff}-i\nu_{n}-\tilde{t}_{{\bf{k}}}\right)b_{{\bf{k}}}(\nu_{n})
\nonumber\\
&&
-\frac{\bar{\Delta}}{\beta{N}}\sum_{{\bf{k}},\nu_{n}}\bar{a}_{{\bf{k}}}(\nu_{n})b_{{\bf{k}}}(\nu_{n})
\nonumber\\
&&
-\frac{{\Delta}}{\beta{N}}\sum_{{\bf{k}},\nu_{n}}\bar{b}_{{\bf{k}}}(\nu_{n})a_{{\bf{k}}}(\nu_{n}),
\label{Equation_35}
\end{eqnarray}
where the effective chemical potentials ${\mu}^{a}_{\rm eff}$ and ${\mu}^{b}_{\rm eff}$ have been introduced as
\begin{eqnarray}
\mu^{a}_{\rm eff}=\epsilon_{a}-\mu+Un_{b}+i\left\langle\dot{\varphi}({{\bf{r}}}\tau)\right\rangle,
\label{Equation_36}
\newline\\
\mu^{b}_{\rm eff}=\epsilon_{b}-\mu+Un_{a}+i\left\langle\dot{\varphi}({{\bf{r}}}\tau)\right\rangle.
\label{Equation_37}
\end{eqnarray}
The factors $n_{a}$ and $n_{b}$ in Eqs.(\ref{Equation_36}) and ({\ref{Equation_37}}) are the average fermion densities $n_{x}=\left\langle n_{x}({\bf{r}}\tau)\right\rangle$. Next, ${t}_{{\bf{k}}}$ and $\tilde{t}_{{\bf{k}}}$ are band-renormalized hopping amplitudes ${t}_{{\bf{k}}}=2tg_{B}\epsilon\left({{\bf{k}}}\right)$ and $\tilde{t}_{{\bf{k}}}=2\tilde{t}g_{B}\epsilon\left({{\bf{k}}}\right)$,
where $g_{B}$ is the bandwidth renormalization factor
\begin{eqnarray}
g_{B}=\left.\left\langle e^{-i[\varphi({{\bf{r}}}\tau)-\varphi({{\bf{r}}}'\tau)]} \right\rangle\right|_{|{\bf{r}}-{\bf{r}}'|={{d}}}.
\label{Equation_38}
\end{eqnarray}
The explicite expression of this important factor will be given in the Section \ref{sec:Section_5}, within the quantum rotor representation. 
$\epsilon\left({{\bf{k}}}\right)$ is the 3D lattice dispersion relation with $d_{\alpha}$ ($\alpha=x,y,z$), being the components of lattice spacing vector ${\bf{d}}={\bf{r}}-{\bf{r}}'$ with ${\bf{r}}$ and ${\bf{r}}'$ n.n. positions
\begin{eqnarray}
\epsilon\left({{\bf{k}}}\right)=\cos(d_{x}k_{x})+\cos(d_{y}k_{y})+\cos(d_{z}k_{z}).
\label{Equation_39}
\end{eqnarray}
For the simple cubic geometry they are all equal: $d_{\alpha}\equiv a$.
Employing the vector-space notations, we can rewrite the action in Eq.(\ref{Equation_35}) in more compact form  
\begin{eqnarray}
{\cal{S}}_{\rm eff}\left[\bar{a},a,\bar{b},b\right]=\frac{1}{\beta{N}}\sum_{{\bf{k}},\nu_{n}}\left[\bar{a}_{\bf{k}}(\nu_{n}),\bar{b}_{\bf{k}}({\nu_{n}})\right]\hat{\cal{G}}^{-1}_{{\bf{k}}}(\nu_{n})\left[\begin{array}{cc}
{a}_{\bf{k}}(\nu_{n})\\
{b}_{\bf{k}}(\nu_{n})
\end{array}
\right]
\nonumber\\
\label{Equation_40}
\end{eqnarray}
with $\hat{\cal{G}}^{-1}_{{\bf{k}}}(\nu_{n})$ inverse of the Green function matrix
\begin{eqnarray}
\hat{\cal{G}}^{-1}_{{\bf{k}}}(\nu_{n})=
\left[
\begin{array}{cc}
{\cal{E}}^{a}_{{\bf{k}}}(\nu_{n})
 & -\bar{\Delta}  \\
-\Delta & {\cal{E}}^{b}_{{\bf{k}}}(\nu_{n})
\end{array}
\right],
\label{Equation_41}
\end{eqnarray}
where the single-particle quasienergies ${\cal{E}}^{a}_{{\bf{k}}}(\nu_{n})$ and ${\cal{E}}^{b}_{{\bf{k}}}(\nu_{n})$ are given after Eq.(\ref{Equation_35})
\begin{eqnarray}
{\cal{E}}^{a}_{{\bf{k}}}(\nu_{n})={\mu}^{a}_{\rm eff}-i\nu_{n}-{t}_{{\bf{k}}},
\label{Equation_42}
\newline\\
{\cal{E}}^{b}_{{\bf{k}}}(\nu_{n})={\mu}^{b}_{\rm eff}-i\nu_{n}-\tilde{t}_{{\bf{k}}}.
\label{Equation_43}
\end{eqnarray}

The general form of the normal fermionic propagator ${\cal{G}}^{\tilde{x}\tilde{x}}({\bf{r}}\tau,{\bf{r}}'\tau')$, defined in terms of the transformed fermionic variables $\tilde{x}=a,b$ is 
\begin{eqnarray}
{\cal{G}}^{\tilde{x}\tilde{x}}({\bf{r}}\tau,{\bf{r}}'\tau')=-\left\langle \tilde{x}({\bf{r}}\tau)\bar{\tilde{x}}({\bf{r}}'\tau')\right\rangle
\label{Equation_44}
\end{eqnarray}
and the anomalous or, the excitonic propagator, is given by
\begin{eqnarray}
{\cal{G}}^{ab}({\bf{r}}\tau,{\bf{r}}'\tau')=\left\langle \bar{a}({\bf{r}}\tau)b({\bf{r}}'\tau')\right\rangle.
\label{Equation_45}
\end{eqnarray}
The averages in Eqs.(\ref{Equation_44}) and (\ref{Equation_45}) are defined with the help of the effective fermionic action in Eq.(\ref{Equation_40})
\begin{eqnarray}
\langle ... \rangle =\frac{\int\left[{\cal{D}}\bar{a}{\cal{D}}a\right]\left[{\cal{D}}\bar{b}{\cal{D}}b\right]...e^{-{\cal{S}}_{\rm eff}\left[\bar{a},a,\bar{b},b\right]}}{\int\left[{\cal{D}}\bar{a}{\cal{D}}a\right]\left[{\cal{D}}\bar{b}{\cal{D}}b\right]e^{-{\cal{S}}_{\rm eff}\left[\bar{a},a,\bar{b},b\right]}}.
\label{Equation_46}
\end{eqnarray}
As a consequence, using Eqs.(\ref{Equation_40}) and (\ref{Equation_46}) we have
\begin{eqnarray}
{\cal{G}}^{aa}({\bf{r}}\tau,{\bf{r}}'\tau')=-\frac{1}{\beta{N}}\sum_{{\bf{k}},\nu_{n}}{\cal{E}}^{a}_{{\bf{k}}}(\nu_{n})\frac{e^{i\left[{\bf{k}}({\bf{r}}-{\bf{r}}')-\nu_{n}(\tau-\tau')\right]}}{{\cal{E}}^{a}_{{\bf{k}}}(\nu_{n}){\cal{E}}^{b}_{{\bf{k}}}(\nu_{n})-|\Delta|^{2}}.
\label{Equation_47}
\nonumber\\
\ \ \
\end{eqnarray}
A similar expression for ${\cal{G}}^{bb}({\bf{r}}\tau,{\bf{r}}'\tau')$ could be obtained with the simple replacement ${\cal{E}}^{a}_{{\bf{k}}}(\nu_{n})\rightarrow {\cal{E}}^{b}_{{\bf{k}}}(\nu_{n})$. Furthermore, for the anomalous propagator we obtain
\begin{eqnarray}
{\cal{G}}^{ab}({\bf{r}}\tau,{\bf{r}}'\tau')=-\frac{\bar{\Delta}}{\beta{N}}\sum_{{\bf{k}},\nu_{n}} \frac{e^{-i\left[{\bf{k}}({\bf{r}}-{\bf{r}}')-\nu_{n}(\tau-\tau')\right]}}{{\cal{E}}^{a}_{{\bf{k}}}(\nu_{n}){\cal{E}}^{b}_{{\bf{k}}}(\nu_{n})-|\Delta|^{2}},
\label{Equation_48}
\nonumber\\
\ \ \ 
\end{eqnarray}
while ${\cal{G}}^{ba}({\bf{r}}\tau,{\bf{r}}'\tau')$ is obtained by the substitution $\bar{\Delta}\rightarrow \Delta$.

\subsection{\label{sec:Section_4_1} Self-consistent solution for $\Delta$, $\Delta_{g}$ and $\Delta_{c}$}
%
Using the local expressions of the Green functions in Eqs.(\ref{Equation_47}) and (\ref{Equation_48}) obtained above, we have the equations for average electron densities $n_{a}$ and $ n_{b}$ corresponding to the $a$ and $b$-orbitals respectively, and also a self-consistent equation for the excitonic order parameter $\Delta$. We have
\begin{center}
\begin{eqnarray}
&&n_{a}={\cal{G}}^{aa}({\bf{0}},0),
\label{Equation_49}
\newline\\
&&n_{b}={\cal{G}}^{bb}({\bf{0}},0),
\label{Equation_50}
\newline\\
&&\Delta=U{\cal{G}}^{ab}({\bf{0}},0).
\label{Equation_51}
\end{eqnarray}
\end{center}
Then, summing over the fermionic Matsubara frequencies, we can rewrite an equivalent system of equations
\begin{eqnarray}
&&\frac{1}{N}\sum_{{\bf{k}}}\left[n_{F}(E^{+}_{{\bf{k}}})+n_{F}(E^{-}_{{\bf{k}}})\right]=1,
\label{Equation_52}
\newline\\
&&\tilde{n}=\frac{1}{N}\sum_{{\bf{k}}}\xi_{{\bf{k}}}\cdot\frac{n_{F}(E^{+}_{{\bf{k}}})-n_{F}(E^{-}_{{\bf{k}}})}{\sqrt{\xi^{2}_{{\bf{k}}}+4\Delta^{2}}},
\label{Equation_53} 
\newline\\
&&\Delta=-\frac{U\Delta}{N}\sum_{{\bf{k}}}\frac{n_{F}(E^{+}_{{\bf{k}}})-n_{F}(E^{-}_{{\bf{k}}})}{\sqrt{\xi^{2}_{{\bf{k}}}+4\Delta^{2}}}.
\label{Equation_54}  
\end{eqnarray}
It is worth to mention that the only difference between the obtained MF-like equations Eqs.(\ref{Equation_52}-\ref{Equation_54}) and the usual HF theory results given in  Ref.\ \onlinecite{Seki} lies in the presence of the bandwidth renormalization factor $g_{B}$ attached to the $c$ and $f$-band's hopping amplitudes $t$ and $\tilde{t}$. In the low-temperature limit this factor goes to $1$ and for $T=0$ $g_{B}=1$ for all values of the Coulomb interaction parameter (see also the discussion at the end of the Section \ref{sec:Section_5_2}). 
%
\begin{figure}[!hbt]
\includegraphics[scale=.8]{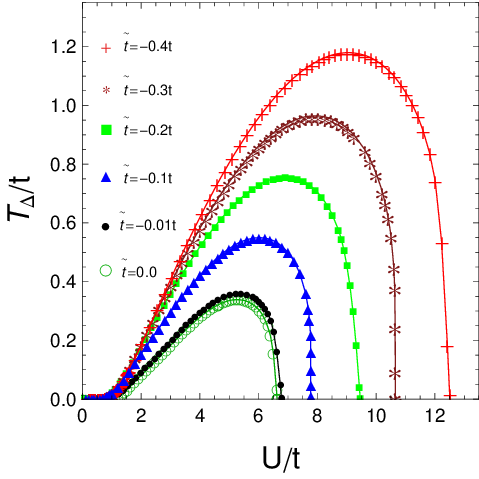}
\caption{\label{fig:Fig_1} The temperature of the excitonic gap formation $T_{\Delta}$ as a function of the interaction parameter $U/t$ for different values of the hopping amplitude $\tilde{t}$. 
}
\end{figure}
%
\begin{figure}[!hbt]
\includegraphics[scale=.8]{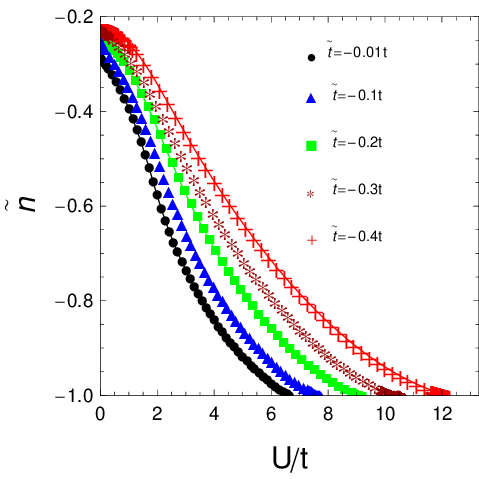}
\caption{\label{fig:Fig_2} The average particle density difference $\tilde{n}$ between the conduction band and valence band, along the pair formation boundary ($\Delta=0$) as a function of the interaction parameter $U/t$. Different values of the hopping amplitude $\tilde{t}$ are considered. 
}
\end{figure}
%
\begin{figure}[!hbt]
\includegraphics[scale=.8]{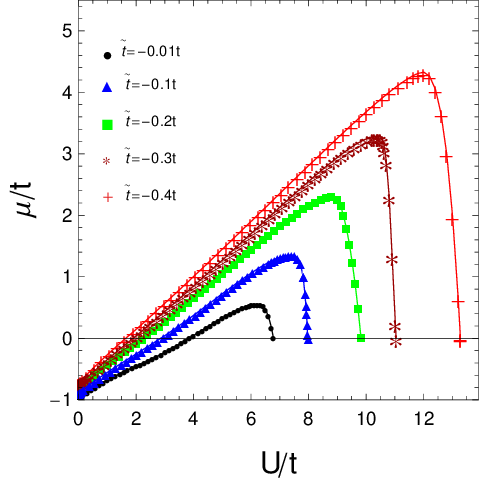}
\caption{\label{fig:Fig_3} The chemical potential $\mu$ at $T\neq 0$ along the pair formation transition boundary ($\Delta=0$) (EI stability region) as a function of the interaction parameter $U/t$. Different values of the hopping amplitude $\tilde{t}$ are considered.}
\end{figure}
%
%
\begin{figure}[!hbt]
\includegraphics[scale=.8]{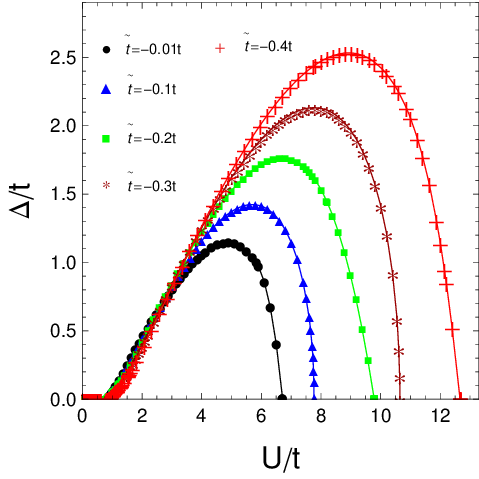}
\caption{\label{fig:Fig_4} The excitonic gap parameter $\Delta$ as a function of the interaction parameter $U/t$ for different values of hopping parameter $\tilde{t}$. The case $T=0$ is considered.}
\end{figure}
%
\begin{figure}[!hbt]
\includegraphics[scale=.8]{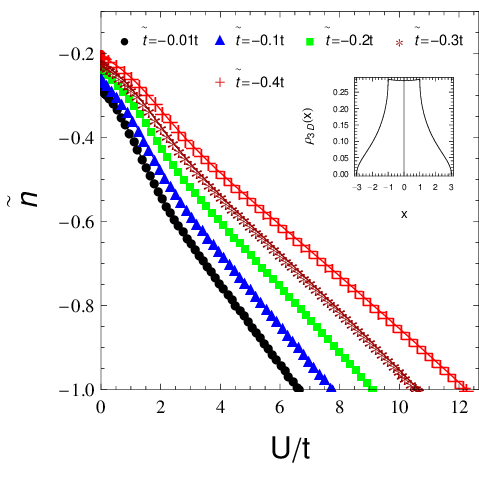}
\caption{\label{fig:Fig_5} The difference $\tilde{n}$ between average electron densities of conduction band and valence band, as a function of the interaction parameter $U/t$ for a number of values of the $f$-band hopping amplitude $\tilde{t}$. The case $T=0$ is considered. The inset shows the density of states (DOS) for the 3D cubic lattice.}
\end{figure}
%
Here, we assumed the half-filled band case ${n}=n_{a}+n_{b}=1$ and we defined the fermion density difference $\tilde{n}=n_{a}-n_{b}$. Without any loss of generality, we have supposed the case of the EI state with the uniform real gap parameter $\Delta$. Furthermore, $n_{F}$ denotes the Fermi-Dirac distribution function $n_{F}(\epsilon)=1/\left(e^{\beta\epsilon}+1\right)$. Next, we have the band-energy parameters $E^{+}_{{\bf{k}}}$ and $E^{-}_{{\bf{k}}}$ defined as
\begin{eqnarray}
&&E^{\pm}_{{\bf{k}}}=\frac{1}{2}\left(-\tilde{t}_{{\bf{k}}}+{\mu}^{b}_{\rm eff}-{t}_{{\bf{k}}}+{\mu}^{a}_{\rm eff}\pm{\sqrt{\xi^{2}_{{\bf{k}}}+4\Delta^{2}}}\right)
\nonumber\\
\label{Equation_55}
\end{eqnarray}
with the quasiparticle dispersion $\xi_{{\bf{k}}}$ 
\begin{eqnarray}
\xi_{{\bf{k}}}=\tilde{t}_{{\bf{k}}}+{\mu}^{a}_{\rm eff}-{t}_{{\bf{k}}}-{\mu}^{b}_{\rm eff}.
\label{Equation_56}
\end{eqnarray}
The energy difference $E^{+}_{{\bf{k}}}-E^{-}_{{\bf{k}}}$
\begin{eqnarray}
\Delta_{c}=E^{+}_{{\bf{k}}}-E^{-}_{{\bf{k}}}
\label{Equation_57}
\end{eqnarray}
defines the \textit{{charge-transfer gap}}, which we will discuss later on in this Section. 

\subsection{\label{sec:Section_4_2} Numerical results and discussion}

The quantities $n_{a}$, $n_{b}$, $\Delta$, $\mu$ and $\Delta_{c}$ can be determined by solving numerically of Eqs.(\ref{Equation_52}) - (\ref{Equation_54}) in a self-consistent way. We start with the discussion of the stability region for the EI phase on the $T-U$ plane, when approaching EI gap to zero: $\Delta \rightarrow 0$. The temperature $T_{\Delta}$ of the excitonic pair formation,  the function $\tilde{n}$ and the chemical potential $\mu$ are considered here.    
The summations over the wave vectors in Eqs.(\ref{Equation_52}-\ref{Equation_54}) can be simplified by introducing the appropriate density of states (DOS) $\rho_{3D}(x)$ for the 3D lattice. Using Eq.(\ref{Equation_39}) we have
\begin{eqnarray}
\rho_{3D}(x)=\frac{1}{N}\sum_{{\bf{k}}}\delta\left[x-\epsilon({\bf{k}})\right].\
\label{Equation_58}
\end{eqnarray}
For the simple cubic lattice the density of states is given as
\begin{eqnarray}
\rho_{3D}(x)=\frac{1}{\pi^{3}}\int^{\min(1,2-x)}_{\max(-1,-2-x)}dy \frac{\Theta\left(1-\frac{|x|}{3}\right)}{\sqrt{1-y^{2}}}
\cdot{{\bf{K}}\left[\sqrt{1-\left(\frac{y}{2}+\frac{x}{2}\right)^{2}}\right]},
\label{Equation_59}
\nonumber\\
\ \ \ 
\end{eqnarray}
where $\Theta(x)$ is the Heaviside step function and ${\bf{K}}(x)$ is the elliptic function of the first kind.\cite{Abramovich} In Fig.~\ref{fig:Fig_1} we have presented the solution for the EI stability region in 3D EFKM by solving the equation $\Delta(T, U)=0$, which determines the temperature $T_{\Delta}$ for which the pairing gap vanishes. The lowest curve in the Fig.~\ref{fig:Fig_1} corresponds to the case of the vanishing narrow-band hopping $\tilde{t}=0$. \cite{Farkasovski_2} In this case the critical temperature $T_{\Delta}$ still finite. Above this temperature, i.e. when $T \gtrsim T_{\Delta}$ we are in the normal Band-Insluator (BI) regime, and $\Delta=0$. Just below the temperature $T_{\Delta}$, i.e. when $T \lesssim T_{\Delta}$, the pair formation began and the system is passing into the EI regime.  
Our calculations, regarding the temperature $T_{\Delta}$ of the pair formation, agree very well with the analogous results in previous works (see Refs.\ \onlinecite{Zenker_1,Zenker_2, Seki, Golosov, Farkasovsky_1, Zenker_3}). For the completeness, the density difference between the conduction band and valence band, and the solutions of chemical potential at the EI transition boundary are presented in Figs.~\ref{fig:Fig_2} and ~\ref{fig:Fig_3}. 

In Fig.~\ref{fig:Fig_4} the solution for the excitonic pairing gap $\Delta$ is plotted as a function of $U/t$ for different values of the $f$-band hopping amplitude $\tilde{t}$ and for $T=0$. The excitonic gap is non-zero for a rather large domain of the Coulomb interaction in agreement with the 3D result of Ref.\ \onlinecite{Zenker_1} and in contrast with the results for the 2D square lattice in Ref.\ \onlinecite{Seki}. The obtained values for the lower and upper bounds of the Coulomb interaction in Ref.\ \onlinecite{Seki} are about $(U_{c1},U_{c2})=(0.66, 6.95)$ and, as it could be expected, they differ considerably from our results, especially for the large hopping.

The solution for $\tilde{n}$ is plotted in Fig.~\ref{fig:Fig_5} as a function of the dimensionless Coulomb interaction parameter $U/t$. It is clear in Fig.~\ref{fig:Fig_5} that in the strong coupling limit $U/t\gg 1$ the system is in the BI regime, because at the upper bound of the Coulomb interaction the $f$-band is fully occupied ($n_{b}=1$) and the $c$-band is totally empty ($n_{a}=0$). In the inset in Fig.~\ref{fig:Fig_5} the plot of the function $\rho_{3D}(x)$ is presented.

\begin{figure}[!hbt]
\includegraphics[scale=.8]{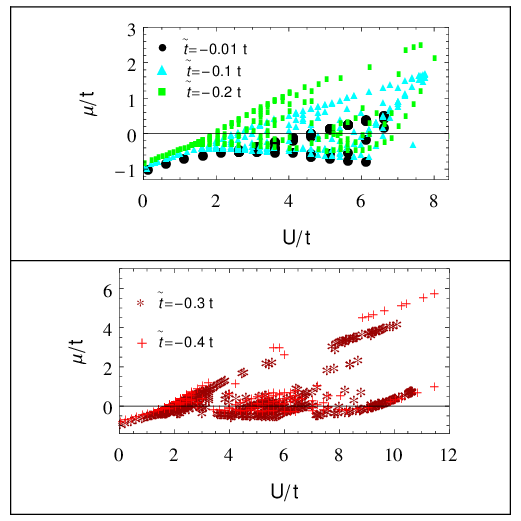}
\caption{\label{fig:Fig_6} The solution for the chemical potential $\mu$ at $T=0$, as a function of the Coulomb interaction parameter $U/t$. Different values of the hopping amplitude $\tilde{t}$ are considered.}
\end{figure}

\begin{figure}[!hbt]
\includegraphics[scale=.8]{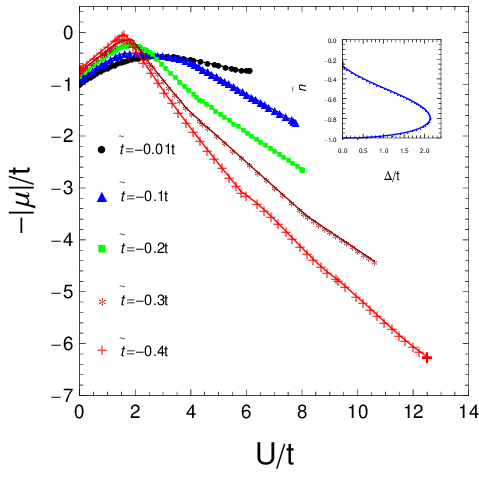}
\caption{\label{fig:Fig_7} The upper bound of the chemical potential $\mu$ accompanying the excitonic pair formation transition as a function of the Coulomb interaction parameter $U/t$ and for different values of the $f$-band hopping amplitude $\tilde{t}$. The case $T=0$ is considered. In the inset, the variation of $\tilde{n}$ is presented as a function of the normalized excitonic gap parameter $\Delta/t$ for $\tilde{t}=-0.3t$ and for $T=0$.}
\end{figure}

The exact numerical solutions for the chemical potential at $T=0$ in the intermediate and strong interaction limits (for example $1.8 \leq U/t \leq 12$ for $\tilde{t}=-0.4t$) form a well defined band (see the leaf-like structures in Fig.~\ref{fig:Fig_6}) for all values of $\tilde{t}$,  and a single particle excitation gap $\Delta_{g}=\mu^{max}-\mu^{min}$ is opening, where $\mu^{max}$ and $\mu^{min}$ are the upper and lower bounds of the chemical potential. The evolution of the upper bound of the chemical potential, as a function of Coulomb interaction parameter $U/t$, is presented in Fig.~\ref{fig:Fig_7}. 

By moving from weak into intermediate coupling regime, the single-particle gap $\Delta_{g}$ and the pairing gap parameter $\Delta$, both are increasing, while in the strong coupling limit ($U/t >8$ for $\tilde{t}=-0.3t$ as an example) $\Delta$ decreases rapidly with increasing $U/t$ while $\Delta_{g}$ remains open (the Hartree-like gap structure).

In the case of vanishing of the pairing gap $\Delta=0$, the single particle gap $\Delta_{g}$ collapses $\Delta_{g}\rightarrow 0$ and the solution for the chemical potential is a single valued (see Fig.~\ref{fig:Fig_3}) (this case corresponds to the case of the boundary of the EI state and is discussed above in Figs.~\ref{fig:Fig_1} - ~\ref{fig:Fig_3}). In other words, we can conclude, that in  case of intermediate and strong Coulomb interaction parameter, the pairing interaction (when $\Delta\neq0$) removes in some sense the degeneracy related to the chemical potential $\mu$. Indeed, the difference between Fig.~\ref{fig:Fig_3} and Fig.~\ref{fig:Fig_6} is due to the pairing interaction $\Delta$. 
%
%

\begin{widetext}
\begin{figure}[htb]
\centering
  \subfloat[\label{fig:Fig_8}{The momentum dependence of the charge-transfer gap $\Delta_{c}$ along the direction $(0,0,0)\rightarrow(\pi,\pi,\pi)$ for $\tilde{t}=-0.3t$ in the extended zone scheme. The wave vector ${\bf{k}}$ is measured in units of $2\pi/d$. The plots are given for different values of the Coulomb energy $U/t$.}]{%
   \centerline{\includegraphics[scale=.8]{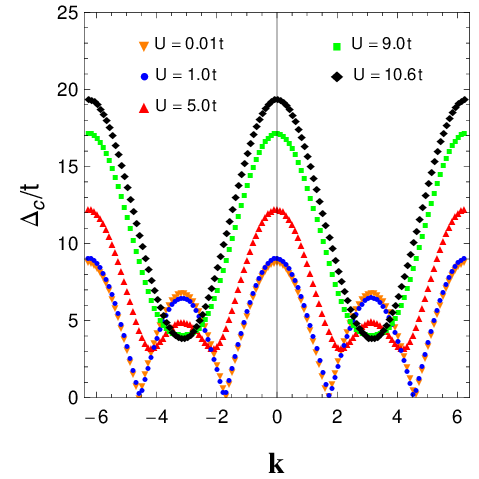}}}
  \subfloat[ \label{fig:Fig_9}{The momentum dependence of the charge-transfer gap $\Delta_{c}$ along the direction $(0,0,0)\rightarrow(\pi,\pi,\pi)$ for $\tilde{t}=-0.1t$ in the extended zone scheme. The wave vector ${\bf{k}}$ is measured in units of $2\pi/d$. The plots are given for different values of the Coulomb energy $U/t$.}]{%
    \centerline{\includegraphics[scale=.8]{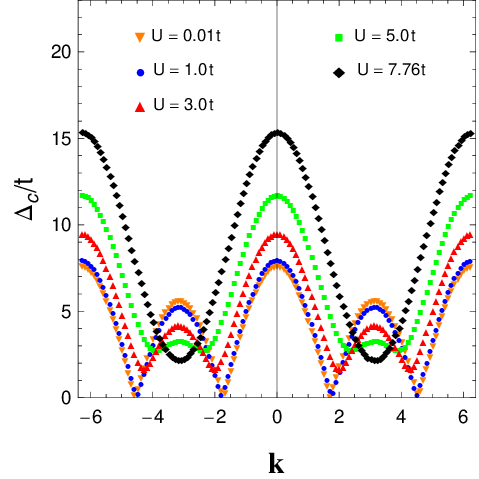}}} 
    \caption{The charge transfer gap $\Delta_{c}$ at $T=0$.}
\end{figure}
\end{widetext}

The charge-transfer gap $\Delta_{c}$ defined in Eq.(\ref{Equation_57}) is calculated as a function of the Coulomb interaction parameter $U/t$. The results are presented in Figs.~\ref{fig:Fig_8} and \ref{fig:Fig_9}.
%
%
We see in Fig.~\ref{fig:Fig_8}, that for the small values of the Coulomb interaction, the charge-transfer gap is nearly zero. The small value of it is the manifestation of the semimetallic limit or the BCS limit. By augmenting the interaction parameter $U$, the gap $\Delta_{c}$ is gradually opening. In Fig.~\ref{fig:Fig_9} we presented the charge-transfer gap for a smaller value of the hopping amplitude $\tilde{t}=-0.1t$. With decreasing the hopping amplitude we are decreasing also the charge-transfer gap. This is consistent with the results for the excitonic gap parameter $\Delta$ presented in Fig.~\ref{fig:Fig_4} and with the behavior of the single particle excitation gap  ($\Delta_{g} $) given in Fig.~\ref{fig:Fig_6}. 
\twocolumngrid
\onecolumngrid
%
\section{\label{sec:Section_5} Quantum rotor representation}
%
\subsection{\label{sec:Section_5_1} Effective phase action}
%
We are interesting now in purely phase action and thus, we will integrate out the fermions in Eq.(\ref{Equation_29}) to obtain the effective phase action in the model. The partition function of the phase-only model is
\begin{eqnarray}
{\cal{Z}}=\int\left[{\cal{D}}\varphi\right]e^{-{\cal{S}}_{\rm eff}\left[\varphi\right]},
\label{Equation_60}
\end{eqnarray}
where 
\begin{eqnarray}
{\cal{S}}_{\rm eff}[\varphi]={\cal{S}}_{0}[\varphi]-\frac{1}{2}\left\langle{\cal{S}}^{2}\right\rangle_{{\cal{S}_{\rm eff}}\left[\bar{a},a,{\bar{b}},b\right]},
\label{Equation_61}
\end{eqnarray}
and the action ${\cal{S}}_{0}[\varphi]$ is given in Eq.(\ref{Equation_28}).
The detailed calculation of the average of second order term in Eq.(\ref{Equation_61}) is given in the Appendix \ref{sec:Section_A}. As result, we have for the phase-only action 
\begin{eqnarray}
&&{\cal{S}}_{{{J}}}\left[\varphi\right]=-\frac{J}{2}\int^{\beta}_{0}d\tau\sum_{\left\langle{\bf{r}},{\bf{r}}'\right\rangle}\cos{2\left[{\varphi}({\bf{r}}\tau)-{\varphi}({\bf{r}}'\tau)\right]},
\nonumber\\
\ \ \
\label{Equation_62}
\end{eqnarray}
where, for the parameter $J$, we have the expresison (see Appendix \ref{sec:Section_A})
\begin{widetext}
\begin{eqnarray}
J=\frac{4\Delta^{2}t\tilde{t}}{9}\int\int{dxdy}\frac{\rho_{3D}(x)\rho_{3D}(y)\epsilon\left(x\right)\epsilon\left(y\right)}{{\sqrt{\xi^{2}(x)+4\Delta^{2}}}}\cdot\left[\Lambda_{1}(x,y)\tanh\left(\frac{\beta E^{+}(x)}{2}\right)-\Lambda_{2}(x,y)\tanh\left(\frac{\beta E^{-}(x)}{2}\right)\right]
\label{Equation_63}
\end{eqnarray}
\end{widetext}
with the parameters $\Lambda_{1}(x,y)$ and $\Lambda_{2}(x,y)$ 
\begin{eqnarray}
\Lambda_{1}(x,y)=\frac{1}{E^{+}(x) - E^{+}(y)}\cdot\frac{1}{E^{+}(x) - E^{-}(y)},
\label{Equation_64}
\newline\\
\Lambda_{2}(x,y)=\frac{1}{E^{-}(x) - E^{-}(y)}\cdot\frac{1}{E^{-}(x) - E^{+}(y)}.
\label{Equation_65}
\nonumber\\
\ \ \ 
\end{eqnarray}
Here $E^{+}(x)$ and $E^{-}(x)$ are the contineous versions (we have just replaced here the index ${\bf{k}}$ in Eq.(\ref{Equation_55}) by the contineous variables $x$) of the similar parameters given in Eq.(\ref{Equation_55}). 
%
\begin{figure}[!hbt]
\includegraphics[scale=.9]{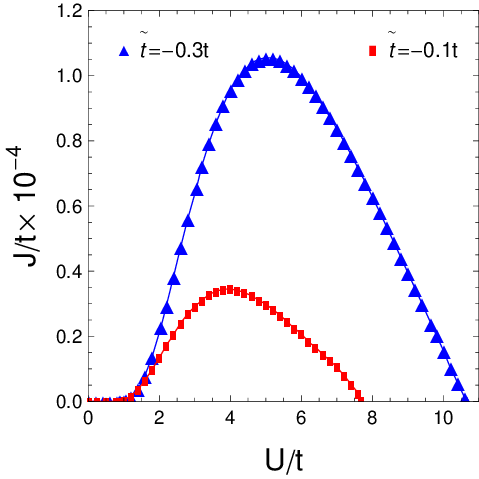}
\caption{\label{fig:Fig_10} The phase-stiffness parameter $J$ at $T=0$ as a function of the interaction parameter $U/t$ and for two different values of the $f$-band hopping amplitude $\tilde{t}$.}
\end{figure}
%
From the expression of the parameter $J$ in Eq.(\ref{Equation_63}) it follows that the non-zero value of this quantity is directly linked with the pairing gap $\Delta$ since $J(\Delta=0)=0$. In Fig.~\ref{fig:Fig_10} we presented the  parameter $J$ as a function of Coulomb interaction parameter $U/t$ and for two different values of the $f$-band hopping $\tilde{t}$. As figures show, the values of $J$ are strictly positive for all regions of the normalized Coulomb interaction parameter. Indeed, the parameter $J$, in units of the hopping parameter $t$ is very small $J/t\ll 1$, but is persistent in the whole interaction region with non-vanishing values of the pairing gap $\Delta$. It is also important to emphasize on the form of the phase-stiffness parameter $J$ in Eq.(\ref{Equation_63}). Especially, it follows from Eq.(\ref{Equation_63}) that the macroscopic phase coherence in the system is characterized by an energy scale $J_{\rm ex}\thicksim (\Delta t_{e}t_{h})/({t_{e}+t_{h}})$ for all values of the Coulomb interaction parameter $U$, which is related to the motion of the center of mass of e-h composed particle, because $(t_{e}t_{h})/(t_{e}+t_{h}) \approx (m_{e}+m_{h})^{-1}$. For the strong interaction limit we are converging with the hard core Boson model, with the kinetic energy proportional to ${\Delta}t_{e}{t}_{h}/U$ ($\Delta$ being the local excitonic order parameter). Thereby, we have shown that non-local correlations between the electrons and holes of different n.n. excitonic pairs, are related with the excitonic BEC condensation. 

In the discussion above, we have derived the effective phase-only action ${\cal{S}}_{\rm eff}\left[\varphi\right]={\cal{S}}_{0}\left[\varphi\right]+{\cal{S}}_{J}\left[\varphi\right]$.
In the following, we cast the ${\cal{S}}_{\rm eff}\left[\varphi\right]$ into the quantum rotor representation. \cite{Kopec_2} To proceed, we replace the phase degrees of freedom with complex unimodular field $z({\bf{r}}\tau)=e^{i\varphi({\bf{r}}\tau)}$ which satisfies the periodic boundary condition $z({\bf{r}}\beta)=z({\bf{r}}0)$. The spherical constraint, imposed on a set of the unimodular variables $z({\bf{r}}\tau)$ is 
\begin{eqnarray}
\frac{1}{N}\sum_{{\bf{r}}}|z({\bf{r}}\tau)|^{2}=1.
\label{Equation_66}
\end{eqnarray}
Now, we introduce new variables $z({\bf{r}}\tau)$ into the partition function in Eq.(\ref{Equation_59}) in a way, consistent with the Faddeev-Popov method \cite{Popov}
\begin{eqnarray}
\int{\cal{D}}\bar{z}{\cal{D}}{z} \delta\left(\frac{}{}\sum_{{\bf{r}}}|z({\bf{r}}\tau)|^{2}-N\frac{}{}\right)\delta\left(z-e^{i{\varphi}({\bf{r}}\tau)}\right)\delta\left(\bar{z}-e^{-i{\varphi}({\bf{r}}\tau)}\right)=1.
\label{Equation_67}
\end{eqnarray}
The spherical constraint in the Eqs.(\ref{Equation_66}) and \ref{Equation_67} can be resolved by introducing the Lagrange multiplier $\lambda$ resulting from the Laplace transform of functional delta representation \cite{Kopec_2}
\begin{eqnarray}
\delta\left(\frac{}{}\sum_{{\bf{r}}}|z({\bf{r}}\tau)|^{2}-N\frac{}{}\right)=\int^{+i\infty}_{-i\infty}\left[\frac{{\cal{D}}\lambda}{2\pi{i}}\right]\times
\nonumber\\
\times e^{-i\int^{\beta}_{0}d\tau\sum_{{\bf{r}}} \lambda\left(\frac{}{}|z({\bf{r}}\tau)|^{2}-1\frac{}{}\right)}.
\label{Equation_68}
\end{eqnarray}
This adds a quadratic term (in the $z$-field) to the phase action. Next, the phase  action in Eq.(\ref{Equation_61}) can be rewritten in more convenient form using the trigonometric half-angle transformation formula 
\begin{eqnarray}
\cos 2\left[{\varphi}({\bf{r}}\tau)-{\varphi}({\bf{r}}'\tau)\right]=2\cos^{2}\left[{\varphi}({\bf{r}}\tau)-{\varphi}({\bf{r}}'\tau)\right]-1.
\nonumber\\
\ \ \
\label{Equation_69}
\end{eqnarray}
Then, in terms of complex variables $z({\bf{r}}\tau)$, the transformation in Eq.(\ref{Equation_69}) leads to a biquadratic term in the phase  action in Eq.(\ref{Equation_61}). We have
\begin{eqnarray}
&&{\cal{S}}_{{{J}}}\left[\varphi\right]\rightarrow{\cal{S}}_{{{J}}}\left[\bar{z},z\right]
\nonumber\\
&&=-\frac{J}{4}\int^{\beta}_{0}d\tau\sum_{\left\langle{\bf{r}},{\bf{r}}'\right\rangle}\left[\bar{z}({\bf{r}}\tau)z({\bf{r}}'\tau)+c.c.\right]^{2}.
\nonumber\\
\ \ \
\label{Equation_70}
\end{eqnarray}
We can rewrite now the partition function in the form
\begin{eqnarray}
{\cal{Z}}=&&\int\left[{\cal{D}}\lambda\right]\left[{\cal{D}}\bar{z}{\cal{D}}z\right]\left[{\cal{D}}\varphi\right]e^{-{\cal{S}}_{0}\left[\varphi\right]}e^{\frac{J}{4}\int^{\beta}_{0}d\tau\sum_{\left\langle{\bf{r}},{\bf{r}}'\right\rangle}\left[\bar{z}\left({\bf{r}}\tau\right)z\left({\bf{r}}'\tau\right)+c.c.\right]^{2}}\times
\nonumber\\
&&\times e^{i\int^{\beta}_{0}d\tau\sum_{{\bf{r}}}\lambda\left(|z({\bf{r}}\tau)|^{2}-1\right)}\delta\left(z-e^{i{\varphi}({\bf{r}}\tau)}\right)\delta\left(\bar{z}-e^{-i{\varphi}({\bf{r}}\tau)}\right).
\label{Equation_71}
\end{eqnarray}
Furthermore, we linearize the action in Eq.(\ref{Equation_70}) (for details see in Appendix \ref{Section_B}) and after we integrate out the phase variables in Eq.(\ref{Equation_59}). Then, after the Fourier transformation of $z$-variables $z({\bf{r}}\tau)=\frac{1}{\beta{N}}\sum_{{\bf{k}},\omega_{n}}z({\bf{k}}\omega_{n})e^{i\left({\bf{k}}{\bf{r}}-\omega_{n}\tau\right)}$ with $\omega_{n}$, being the Bose-Matsubara frequencies $\omega_{n}=\frac{2\pi{n}}{\beta}$ with ($n=0,\pm 1,\pm 2,...$), the partition function assumes the form
\begin{eqnarray}
{\cal{Z}}=\int\left[{\cal{D}}\lambda\right]\left[{\cal{D}}\bar{z}{\cal{D}}{z}\right]e^{-{\cal{S}}_{\lambda}[\bar{z},z]},
\label{Equation_72}
\end{eqnarray}
with the action ${\cal{S}}_{\lambda}[\bar{z},z]$
\begin{eqnarray}
{\cal{S}}_{\lambda}[\bar{z},z]&&=\frac{1}{\beta{N}}\sum_{{\bf{k}}\omega_{n}}\bar{z}({\bf{k}}\omega_{n}){\cal{G}}^{-1}_{z}({\bf{k}}\omega_{n})z({\bf{k}}\omega_{n}),
\nonumber\\
\label{Equation_73}
\end{eqnarray}
where
\begin{eqnarray}
{\cal{G}}^{-1}_{z}({\bf{k}}\omega_{n})=\gamma^{-1}(\omega_{n})-4g_{B}J\epsilon({\bf{k}})-\lambda.
\label{Equation_74}
\end{eqnarray}
Furthermore, $g_{b}$ stands for the bandwidth-renormalization factor, $\gamma^{-1}(\omega_{n})$ in Eq.(\ref{Equation_74}) is the inverse of the Fourier transformed two-point phase correlation function $\gamma({\bf{r}}\tau,{\bf{r}}'\tau')$
\begin{eqnarray}
\gamma({\bf{r}}\tau,{\bf{r}}'\tau')=\frac{1}{{\cal{Z}}_{0}}\int \left[{\cal{D}}{\varphi}\right]e^{-{\cal{S}}_{0}[{\varphi}]}e^{i\left[{\varphi}({\bf{r}}\tau)-{\varphi}({\bf{r}}'\tau')\right]},
\label{Equation_75}
\end{eqnarray}
where ${\cal{Z}}_{0}$ is the statistical sum of the noninteracting set of quantum rotators%
\begin{eqnarray}
{\cal{Z}}_{0}=\int\left[{\cal{D}}\varphi \right] e^{-{\cal{S}}_{0}[\varphi]}.
\label{Equation_76}
\end{eqnarray}
The calculation of the Fourier transformation $\gamma(\omega_{n})$ of the function in Eq.(\ref{Equation_75}) is straightforward \cite{Kopec_2}
\begin{eqnarray}
\gamma(\omega_{n})=\frac{8}{U{\cal{Z}}_{0}}\sum^{+\infty}_{{m}=-\infty}\frac{e^{-\frac{U\beta}{4}\left({{m}}-\frac{2\bar{\mu}}{U}\right)^{2}}}{1-16\left[\frac{i\omega_{n}}{U}-\frac{1}{2}\left({{m}}-\frac{2\bar{\mu}}{U}\right)\right]^{2}},
\nonumber\\
\ \ \
\label{Equation_77} 
\end{eqnarray}
where
\begin{eqnarray}
{\cal{Z}}_{0}=\sum^{+\infty}_{{m}=-\infty}e^{-\frac{U\beta}{4}\left({{m}}-\frac{2\bar{\mu}}{U}\right)^{2}}.
\label{Equation_78}
\end{eqnarray}
The summations in Eqs.(\ref{Equation_77}) and (\ref{Equation_78}) are over the winding numbers $m$ of the U(1) group (see the Section \ref{sec:Section_3_1}).
%
\subsection{\label{sec:Section_5_2} Exciton condensate at $T\thicksim T_{c}$}
%
In the thermodynamic limit $N\rightarrow \infty$ the integration over $\lambda$-field in Eq.(\ref{Equation_71}) can be performed exactly using the saddle-point method
\begin{eqnarray}                  
\left. \frac{\delta {\cal{S}}_{\lambda}[\bar{z},z]}{\delta \lambda}\right|_{\lambda=\lambda_{0}}=0.
\label{Equation_79}
\end{eqnarray}
As a result, one can write the constraint for the saddle-point value of the Lagrange multiplier $\lambda_{0}$
\begin{eqnarray}
1=\lim_{\delta \rightarrow 0^{+}} \langle z({\bf{r}}\tau)\bar{z}({\bf{r}}\tau+\delta),
\label{Equation_80}
\end{eqnarray}
where the average in Eq.(\ref{Equation_80}) is defined as
\begin{eqnarray}
\langle ... \rangle\equiv\frac{\int\left[{\cal{D}}\bar{z}{\cal{D}}{z}\right] ... e^{-{\cal{S}}_{\lambda_{0}}[\bar{z},z]}}{\int\left[{\cal{D}}\bar{z}{\cal{D}}{z}\right]e^{-{\cal{S}}_{\lambda_{0}}[\bar{z},z]}}.
\label{Equation_81}
\end{eqnarray}
Then, with the help of the Eq.(\ref{Equation_74}) we can write
\begin{eqnarray}
1=\frac{1}{(\beta{N})^{2}}\sum_{{\bf{k}},,\omega_{n}}\langle z({\bf{k}}\omega_{n})\bar{z}({\bf{k}}\omega_{n})\rangle \equiv \frac{1}{\beta{N}}\sum_{{\bf{k}},\omega_{n}}{\cal{G}}_{z}({\bf{k}}\omega_{n}).
\nonumber\\
\label{Equation_82}
\end{eqnarray}
After Eq.(\ref{Equation_75}), the equation Eq.(\ref{Equation_82}) takes now the following explicit form
\begin{eqnarray}
1=\frac{1}{\beta{N}}\sum_{\substack{{\bf{k}},\omega_{n}}}\frac{1}{\gamma^{-1}(\omega_{n})-4g_{B}J\epsilon({\bf{k}})-\lambda_{0}}.
\nonumber\\
\label{Equation_83}
\end{eqnarray}
The explicite value of the parameter $\lambda_{0}$ could be determined with the help of the Thouless criterion.\cite{Thouless} It states that the uniform static order parameter susceptibility diverges at the phase transition. Thus ${\cal{G}}^{-1}_{z}({\bf{k}}={{0}},\omega_{n}=0)=0$ from which we can derive the critical value of the Lagrange multiplier
\begin{eqnarray}
\gamma^{-1}(\omega_{n}=0)-4g_{B}J\epsilon({\bf{0}})-\lambda_{0}=0.
\label{Equation_84}
\end{eqnarray}
Furthermore, we find 
\begin{eqnarray}
\lambda_{0}=\frac{U}{8}-\frac{2\bar{\mu}^{2}}{U}-4g_{B}J\epsilon({\bf{0}}).
\label{Equation_85}
\end{eqnarray}
After performing the Bose-Matsubara frequency summations in Eq.(\ref{Equation_82}), we obtain the equation for the excitonic BEC transition critical temperature $T_{c}$
\begin{eqnarray}
\frac{U}{4N}\sum_{{\bf{k}}} \frac{n_{B}\left(\zeta_{1{\bf{k}}}\right)-n_{B}\left(\zeta_{2{\bf{k}}}\right)}{\sqrt{\bar{\mu}^{2}+2Ug_{B}{J}\left[\epsilon({{\bf{0}}})-\epsilon({{\bf{k}}})\right]}}=1,
\nonumber\\
\ \ \
\label{Equation_86}
\end{eqnarray}
where $n_{B}\left(\epsilon\right)$ is the Bose-Einstein distribution function $n_{B}\left(\epsilon\right)=1/\left(e^{\beta\epsilon}-1\right)$ and the variables $\zeta_{1{\bf{k}}}$ and $\zeta_{2{\bf{k}}}$ are given by
\begin{eqnarray}
\zeta_{\alpha{\bf{k}}}=-\bar{\mu}-(-1)^{\alpha}{\sqrt{\bar{\mu}^{2}+2U{J}g_{B}\left[\epsilon({{\bf{0}}})-\epsilon({{\bf{k}}})\right]}},
\label{Equation_87}
\end{eqnarray}
%
\begin{figure}[!hbt]
\includegraphics[scale=.8]{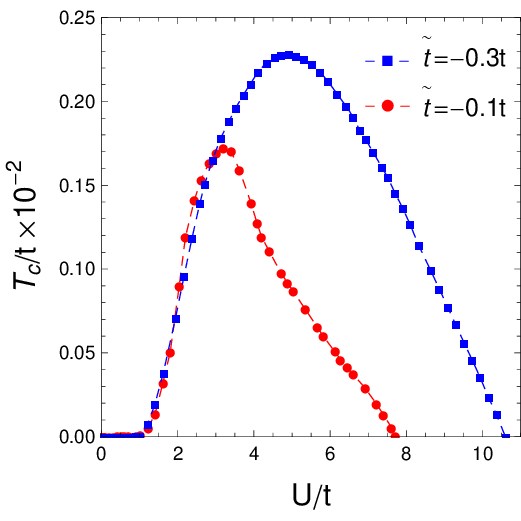}
\caption{\label{fig:Fig_11} The critical temperature $T_{c}/t$ of the excitonic condensate formation as 
a function of the Coulomb interaction parameter $U/t$ and for two different values of the $f$-band hopping amplitude $\tilde{t}$.}
\end{figure}
%
where $\alpha=1,2$. We see also, that at the fundamental state with ${\bf{k}}=0$ there is a residual gap $\Delta W=\zeta_{1}\left({\bf{0}}\right)-\zeta_{2}\left({\bf{0}}\right)=-2\bar{\mu}$ related to the condensate, which equals the binding energy of a molecule in the BEC limit $E_{bind}\approx |2\bar{\mu}|$. \cite{Nozieres, Tsuchiya_1,Tsuchiya_2} 

At the end of this section we give also the analytical expression of the bandwidth-renormalization factor $g_{B}\left({\bf{r}}-{\bf{r}}'\right)$
\begin{eqnarray}
g_{B}\left({\bf{r}}-{\bf{r}}'\right)=\frac{U}{12N}\sum_{{\bf{k}}} \epsilon\left({\bf{k}}\right)\cdot\frac{n_{B}\left(\zeta_{1{\bf{k}}}\right)-n_{B}\left(\zeta_{2{\bf{k}}}\right)}{\sqrt{\bar{\mu}^{2}+2Ug_{B}{J}\left[\epsilon({{\bf{0}}})-\epsilon({{\bf{k}}})\right]}}.
\nonumber\\
\ \ \
\label{Equation_88}
\end{eqnarray}
In fact, the calculation of the factor $g_{B}\left({\bf{r}}-{\bf{r}}'\right)$ could be done alternatively within the self-consistent-harmonic-approximation (SCHA).\cite{Kleinert,Kim} In this approximation the quantum rotor description is reduced to classical Hamiltonian one. We do not present here the SCHA results for $g_{B}\left({\bf{r}}-{\bf{r}}'\right)$. This could be a subject of a future investigation. The calculation of the factor $g_{B}\left({\bf{r}}-{\bf{r}}'\right)$ shows that, at $T=0$, it is equal identically to $1$.

The numerical solution of the equation Eq.(\ref{Equation_86}) is presented in Fig.~\ref{fig:Fig_11}. We see that the excitonic BEC transition critical temperature $T_{c}$ is much smaller than the EP formation critical temperature $T_{\Delta}$ discussed in the Section \ref{sec:Section_4}. This conclusion is in the well agreement with the previous theoretical investigations. \cite{Snoke_1, Snoke_2, Tomio, Micnas}
%
\subsection{BEC transition amplitude at $T\lesssim  T_{c}$}
%
In general case, the local constraint in Eq.(\ref{Equation_66}) for bosonic unimodular variables $z({\bf{r}}\tau)$ breaks down at very low temperatures, (especially at $T=0$) because we have to consider the symmetry breaking related to the Bosonic sector, thus, critically, we have the fluctuation form $z({\bf{r}}\tau)=\left\langle e^{i\varphi}({\bf{r}}\tau)\right\rangle+\tilde{z}({\bf{r}}\tau)$ and the unimodularity constraint is broken.
In the limite of very low temperatures, considering the BEC of excitons, we have the spontaneous breaking of local U(1) gauge-symmetry related to the phase field, leading to the nonvanishing expectation value of $z({\bf{r}}\tau)$. In order to demonstrate this, we separate the single particle state ${\bf{k}}=0$ by using Bogoliubov displacement operation (see, for details in Refs.\ \onlinecite{Moskalenko} and \onlinecite{Simanek}). Then, we write for the complex variables $z({\bf{k}},\omega_{n})$
\begin{eqnarray}
z({\bf{k}},\omega_{n})=\beta{N}\psi_{0}\delta_{{\bf{k}},0}\delta_{\omega_{n},0}+\tilde{z}({\bf{k}},\omega_{n})(1-\delta_{{\bf{k}},0})(1-\delta_{\omega_{n},0}),
\label{Equation_89}
\end{eqnarray}
where $\psi_{0}$ is the condensate transition amplitude $\psi_{0}=\left\langle {z}({\bf{k}},\omega_{n})\right\rangle$ of the bosonic field. Next $\tilde{z}({\bf{k}},\omega_{n})$ is the excitation part of effective Bose-field.
The gerenal form of the bosonic charge propagator is given by
\begin{eqnarray}
{\cal{G}}_{z}({\bf{r}}\tau,{\bf{r}}'\tau')=\left\langle z({\bf{r}}\tau)\bar{z}({\bf{r}}'\tau')\right\rangle=\frac{1}{\beta{N}}\sum_{{\bf{k}},\omega_{n}}{\cal{G}}_{z}({\bf{k}},\omega_{n}) e^{-i\left[{\bf{k}}{\bf{d}}-\omega_{n}\delta\right]},
\label{Equation_90}
\end{eqnarray}
where
\begin{eqnarray}
{\cal{G}}_{z}({\bf{k}},\omega_{n})=\left\langle z({\bf{k}},\omega_{n})\bar{z}({\bf{k}},\omega_{n})\right\rangle 
\label{Equation_91}
\end{eqnarray}
The average in Eq.(\ref{Equation_91}) is defined in Eq.(\ref{Equation_81}).
We consider the expectation value $\left\langle z({\bf{k}},\omega_{n})\bar{z}({\bf{k}},\omega_{n})\right\rangle$ and we draw the condensate part by applying the transformation in Eq.(\ref{Equation_89}). Hence, we have  
\begin{eqnarray}
&&{\cal{G}}_{z}({\bf{k}},\omega_{n})=\left\langle z({\bf{k}},\omega_{n})\bar{z}({\bf{k}},\omega_{n})\right\rangle=
\nonumber\\
&&=\beta{N}|\psi_{0}|^{2}\delta_{{\bf{k}},0}\delta_{\omega_{n},0}+\tilde G_{z}({\bf{k}},\omega_{n}).
\label{Equation_92}
\end{eqnarray}
Here $\tilde G_{z}({\bf{k}},\omega_{n})$ is related to the on-condensate exctitation part of the bosonic sector
\begin{eqnarray}
\tilde G_{z}({\bf{k}},\omega_{n})=\left\langle \tilde{z}({\bf{k}},\omega_{n})\bar{\tilde{z}}({\bf{k}},\omega_{n})\right\rangle
\label{Equation_93}
\end{eqnarray}
%
\begin{figure}[!hbt]
\includegraphics[scale=.8]{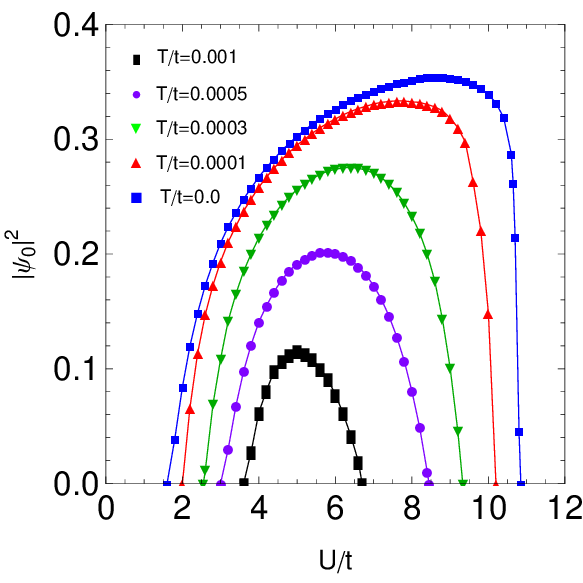}
\caption{\label{fig:Fig_12} The excitonic condensate transition probability function $\psi^{2}_{0}$ in dependence on the normalized Coulomb interaction parameter $U/t$. Different values of temperature are considered.}
\end{figure}
%
Finally, the local constraint in Eq.(\ref{Equation_66}) will be rewritten as
\begin{eqnarray}
1-|\psi_{0}|^{2}=\frac{1}{\beta{N}}\sum_{\substack{{\bf{k}}\neq 0 \\ \omega_{n}\neq 0}}G_{z}({\bf{k}},\omega_{n}),
\label{Equation_94}
\end{eqnarray}
then, we get
\begin{eqnarray}
|\psi_{0}|^{2}=1-\frac{U}{4N}\sum_{{\bf{k}}} \frac{n_{B}\left(\zeta_{1{\bf{k}}}\right)-n_{B}\left(\zeta_{2{\bf{k}}}\right)}{\sqrt{\bar{\mu}^{2}+2Ug_{B}{J}\left[\epsilon({{\bf{0}}})-\epsilon({{\bf{k}}})\right]}}.
\nonumber\\
\label{Equation_95}
\end{eqnarray}
The obtained values for $|\psi_{0}|^{2}$ are plotted in Fig.~\ref{fig:Fig_12}. With increasing the temperature, BEC transition probability decreases and disappears for the high temperature limit. 
%
\section{\label{sec:Section_6} Momentum distribution functions and exciton coherence length}
%
To proceed we define frequency-summed normal and anomalous momentum dependent functions 

\begin{figure}[!hbt]
\includegraphics[scale=.2]{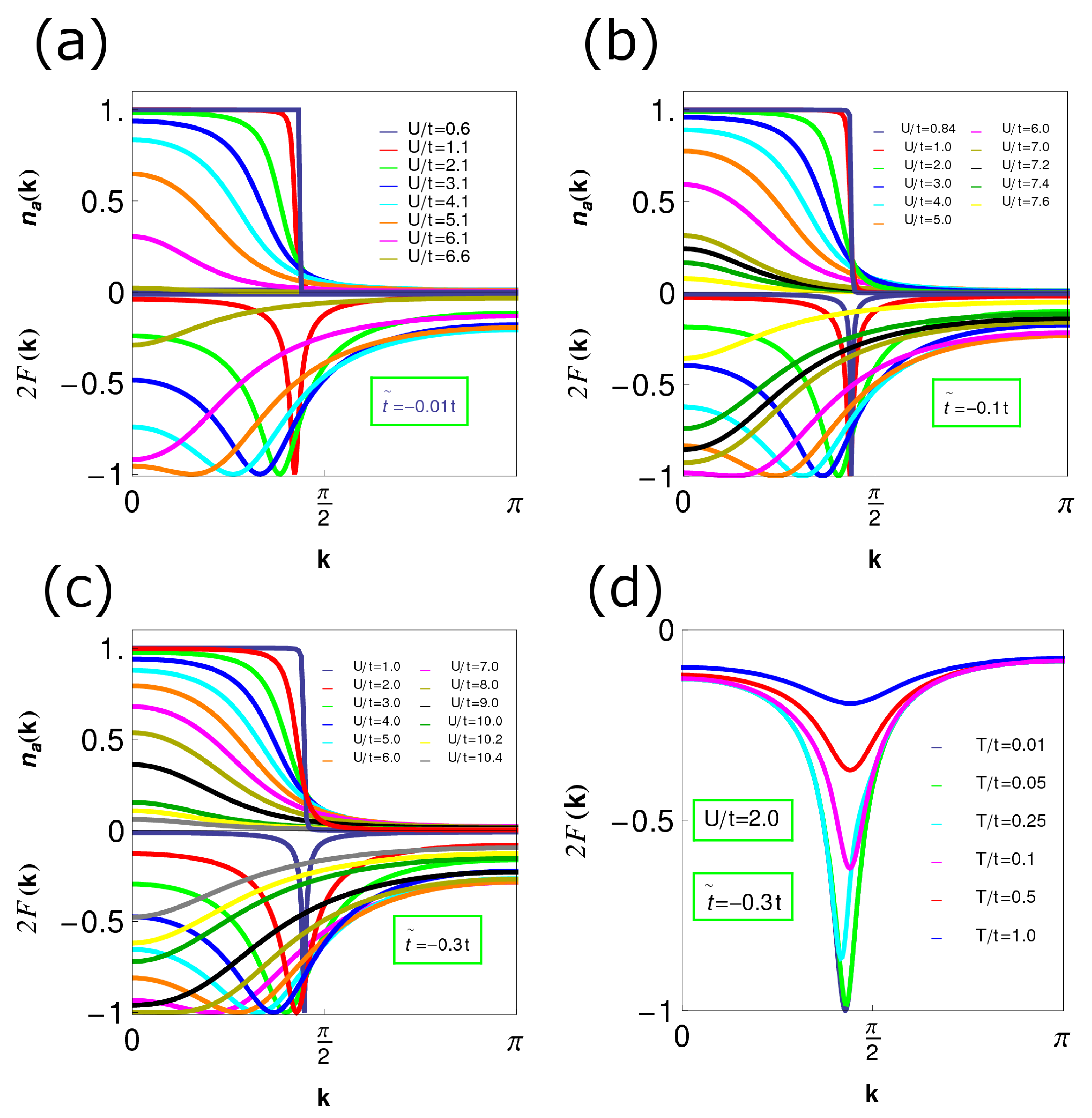}
\caption{\label{fig:Fig_13} \scriptsize{Single particle normal and anomalous momentum distribution functions at $T=0$. (a). The normal momentum distribution function $n_{a}({\bf{k}})$ and anomalous function $2F({\bf{k}})$ along the direction $(0,0,0)\rightarrow(\pi,\pi,\pi)$ at $T=0$ for different values of the normalized Coulomb interaction parameter $U/t$ and for and for $\tilde{t}=-0.01t$. The wave vector ${\bf{k}}$ is given in units of $2\pi/d$. (b) The normal momentum distribution function $n_{a}({\bf{k}})$ and anomalous function $2F({\bf{k}})$ along the direction $(0,0,0)\rightarrow(\pi,\pi,\pi)$ at $T=0$ for different values
of the normalized Coulomb interaction parameter $U/t$ and for and for $\tilde{t}=-0.1t$. The wave vector ${\bf{k}}$ is given in units of $2\pi/d$.(c) The normal momentum distribution function $n_{a}({\bf{k}})$ 
and anomalous function $2F({\bf{k}})$ along the direction $(0,0,0)\rightarrow(\pi,\pi,\pi)$ at $T=0$ and for different values of the normalized Coulomb interaction parameter $U/t$ and for and for $\tilde{t}=-0.3t$. The wave vector ${\bf{k}}$ is given in units of $2\pi/d$. (d) The temperature dependence of the anomalous momentum distribution function $2F({\bf{k}})$ along $(0,0,0)\rightarrow(\pi,\pi,\pi)$ for
$U/t = 2.0$ and for $\tilde{t}=-0.3t$. The wave vector ${\bf{k}}$ is given in units of $2\pi/d$. (Color figure online)}}
\end{figure}

\onecolumngrid

\begin{eqnarray}
n_{a}({{\bf{k}}})=\frac{1}{\beta}\sum_{{\nu_{n}}}{\cal{G}}^{aa}({\bf{k}}\nu_{n}),
\label{Equation_96}
\nonumber\\
F({{\bf{k}}})=\frac{1}{\beta}\sum_{{\nu_{n}}}{\cal{G}}^{ab}({\bf{k}}\nu_{n}),
\label{Equation_97}
\end{eqnarray}
where ${\cal{G}}^{aa}({\bf{k}}\nu_{n})$ and ${\cal{G}}^{ab}({\bf{k}}\nu_{n})$ are the Fourier transformations of the local, normal and anomalous, propagators. Using Eqs.(\ref{Equation_47}) and (\ref{Equation_48}) we obtain
\begin{eqnarray}
n_{a}({{\bf{k}}})=\frac{1}{\beta}\sum_{{\nu_{n}}} \frac{{\mu^{b}_{\rm eff}}-i\nu_{n}-\tilde{t}}{{\cal{E}}^{a}_{{\bf{k}}}(\nu_{n}){\cal{E}}^{b}_{{\bf{k}}}(\nu_{n})-\Delta^{2}},
\label{Equation_98}
\newline\\
F({{\bf{k}}})=-\frac{1}{\beta}\sum_{{\nu_{n}}} \frac{{\Delta}}{{\cal{E}}^{a}_{{\bf{k}}}(\nu_{n}){\cal{E}}^{b}_{{\bf{k}}}(\nu_{n})-\Delta^{2}}.
\label{Equation_99}
\end{eqnarray}
Summing over the fermionic Matsubara frequencies $\nu_{n}$, we get 
\begin{eqnarray}
n_{a}({\bf{k}})=v^{2}_{{\bf{k}}}n_{F}(E^{-}_{{\bf{k}}})-u^{2}_{{\bf{k}}}n_{F}(E^{+}_{{\bf{k}}}),
\label{Equation_100}
\newline\\
F({{\bf{k}}})=u_{{\bf{k}}}v_{{\bf{k}}}\left[n_{F}\left(E^{+}_{{\bf{k}}}\right)-n_{F}\left(E^{-}_{{\bf{k}}}\right)\right],
\label{Equation_101}
\end{eqnarray}
while the function $n_{b}({\bf{k}})$ for the $b$-orbital is simply $n_{b}({\bf{k}})=1-n_{a}({\bf{k}})$. The Bogoliubov coefficients appearing in Eqs.(\ref{Equation_100}) and (\ref{Equation_101}) are given by
\begin{eqnarray}
&&u^{2}_{{\bf{k}}}=\frac{1}{2}\left(1+\frac{\xi_{{\bf{k}}}}{\sqrt{\xi^{2}_{{\bf{k}}}+4\Delta^{2}}}\right),
\label{Equation_102}
\newline\\
&&v^{2}_{{\bf{k}}}=\frac{1}{2}\left (1-\frac{\xi_{{\bf{k}}}}{\sqrt{\xi^{2}_{{\bf{k}}}+4\Delta^{2}}}\right),
\label{Equation_103}
\newline\\
&&u_{{\bf{k}}}v_{{\bf{k}}}=\frac{\Delta}{\sqrt{\xi^{2}_{{\bf{k}}}+4\Delta^{2}}}.
\label{Equation_104}
\end{eqnarray}

%
\twocolumngrid
\begin{widetext}
\begin{figure}[htb]
\centering
  \subfloat[\label{fig:Fig_14}{Coherence length $\xi_{c}$ as a function of the Coulomb interaction parameter $U/t$ in units of lattice constant
 $d$ and at $T=0$. Two different values 
of the hopping amplitude $\tilde{t}$ 
are considered: $\tilde{t}=-0.01t$ and $\tilde{t}=-0.1t$.}]{%
   \centerline{\includegraphics[scale=.8]{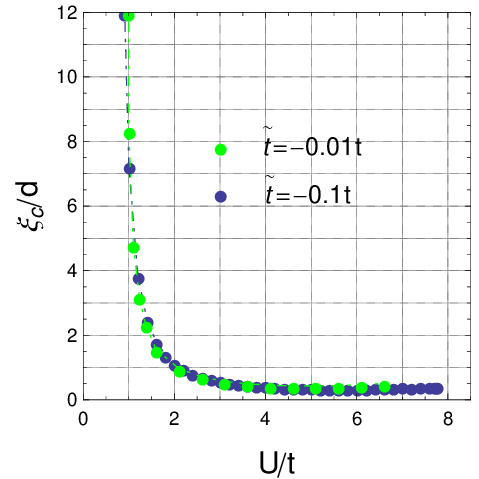}}}
  \subfloat[ \label{fig:Fig_15}{Coherence length $\xi_{c}$ 
as a function of the Coulomb interaction parameter $U/t$ 
in units of the lattice constant $d$ and at $T=0$. 
Two different values of the hopping amplitude 
$\tilde{t}$ are considered: $\tilde{t}=-0.2t$
 and $\tilde{t}=-0.3t$.}]{%
    \centerline{\includegraphics[scale=.8]{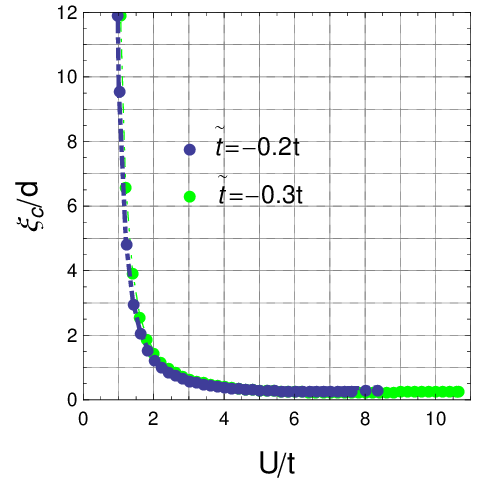}}} 
    \caption{Single exciton coherence length at $T=0$.}
\end{figure}
\end{widetext}
%
\onecolumngrid
The plots of the normal and anomalous functions $n_{a}({\bf{k}})$ and $F({\bf{k}})$ are given in Fig.~\ref{fig:Fig_13} (a)-(d). The ${\bf{k}}$ -summations in the analytical expression of $\xi_{c}$ were done with the $\left(100\times100\times100\right)$ ${\bf{k}}$-points in the FBZ. In the weak coupling regime the normal distribution function $n_{a}({\bf{k}})$ drops at ${\bf{k}}_{F}$ (see the top plots in the panels (a)-(c) in Fig.~\ref{fig:Fig_13}) and anomalous momentum function (the bottom plots in the panels (a)-(c) in Figs.~\ref{fig:Fig_13}) is picked at the Fermi level. With increasing the Coulomb interaction $n_{a}({\bf{k}})$ spread out in the ${\bf{k}}$-space and also ${\bf{k}}_{F}$ becomes broad with the Fermi level $k_{F}$ displaced to the value $(0,0,0)$ in the momentum space. Across the crossover regime, the anomalous momentum function decreases for all momenta of the reciprocal space and this is consistent with the behavior of the excitonic gap parameter $\Delta$ in the strong coupling regime presented in Fig.~\ref{fig:Fig_2}. Subsequently, in the panel in the panels (c) in Fig.~\ref{fig:Fig_13}, we have presented the temperature dependence of the anomalous distribution function. 

The spatial coherence of a fermionic system is encoded in its one-body density matrix, therefore, the anomalous momentum function is directly related to the excitonic coherence length. We can associate a characteristic decay of $F({\bf{k}})$ with the coherence length $\xi_{c}$, defined by the relation \cite{Seki}
\begin{eqnarray}
\xi^{2}_{c}=\frac{\sum_{{\bf{k}}}|\nabla_{{\bf{k}}}{{F}}({\bf{k}})|^{2}}{\sum_{{\bf{k}}}|{{F}}({\bf{k}})|^{2}}.
\label{Equation_105}
\end{eqnarray}
%
The quantity $\xi_{c}$ provides the quantitative information about the properties of the system.
By calculating the coherence length given by Eq.(\ref{Equation_105}) for different values of the Coulomb interaction parameter $U/t$, we can see directly the spatial extension of a single exciton. The results are given in Figs.~\ref{fig:Fig_14} and \ref{fig:Fig_15}, where a rapid growth of the coherence length, for the small values of the Coulomb interaction parameter, is anticipated with the excitons cooled down below the temperature of their quantum degeneracy and the system is in the macroscopic phase-coherent regime. On the other hand, opposite to this behavior, the coherence length decreases rapidly with increasing $U/t$. A very similar decrease of the coherence length of the Cooper pairs is also proved in exact-diagonalization study on the attractive Hubbard model. \cite{Nakauchi} The coherence length has a minimum in the intermediate coupling regime and this is due to the denominator $\sum_{{\bf{k}}}|{{F}}({\bf{k}})|^{2}$ in Eq.(\ref{Equation_105}), which is largest in this case. In the strong interaction region, $\xi_{c}$ slightly increases with increasing $U/t$. Our resutls are in good agreement with the HF results discussed earlier. \cite{Kaneko,Seki,Zenker_3}
%
\section{\label{sec:Section_7} Summary and outlook}

Now it is interesting to relate results of our calculations on the 3D excitonic system to the experimental results, e.g., for the compound TmSe$_{0.45}$Te$_{0.55}$, which is an intermediate valent semiconductor. \cite{Neuenschwander, Wachter, Wachter_1,Wachter_2} The hopping parameter $\tilde{t}$ is estimated for $|\tilde{t}|=0.3|t|=5$ meV (see Ref. \ \onlinecite{Zenker_3}). By using these values, we find for the maximum of the excitonic pair transition temperature $T^{\rm max}_{\Delta}=186.6$ K at $U=8|t|$, while the maximum of the exciton BEC transition temperature is found to be smaller of about two orders of magnitude at $T^{\rm max}_{c}=0.44$ K for $U=4.8|t|$.

Furthermore, the charge-gap bandwidth was found as to be $W=|\Delta^{\rm min}_{c}|=0.0682$ eV and the single particle excitation gap is of order $\Delta_{g}=0.057$ eV at $U=10.6|t|$. The obtained values fit into the experimental results on TmSe$_{0.45}$Te$_{0.55}$, where $T_{\Delta}$ is found to be of order $250$ K and below.\cite{Wachter_2}
For the maximum energy scale corresponding to phase-stiffness parameter $J$  (for $\tilde{t}=-0.3$) we find, for TmSe$_{0.45}$Te$_{0.55}$ $J\approx 0.00176 $ meV (corresponding to $U=0.0833$ eV) or, in temperature units $J\approx 20$ mK. 

In conclusion, we have studied the excitonic phase transition in a system of conduction band electrons with transfer parameter $t$, and the valence band holes, described by 3D extended Falicov-Kimball model with the tunable Coulomb interaction $U$ between both species. To this end we implement the functional integral formulation of our model, where the Coulomb interaction term is expressed in terms of U(1) quantum phase variables $\varphi$ conjugated to the local particle number, providing a useful representation of strongly correlated systems. At low temperatures, the electron-hole system may become unstable with respect to the formation of the excitons at  $T=T_{\Delta}$, exhibiting a gap $\Delta$ in the particle excitation spectrum controlled by the parameter $U/t$, which gives the relevant energy scale for the excitonic insulator state. In the weak coupling limit, $U/t\ll 1$, the binding energy of excitonic pairs is small, thus pair breaking effect controls the excitonic phase transition in analogy to that what happens in a standard BCS superconductor. In the excitonic system with the strong pairing  $U/t\gg 1$, we have the situation, where the pairs are strongly bound and localized which diminish, $T_{\Delta}$ for large $U/t$.

We have shown that the excitonic BEC transition temperature $T_{c}$ is much smaller than the critical temperature $T_{\Delta}$ of the excitonic pair formation in well agreement with the discussions in Refs.\ \onlinecite{Snoke_1}, \onlinecite{Snoke_2} and \onlinecite{Tomio}. Our results are in good agreement with the previous theoretical and experimental results. A possible direction for future work will be the determination of the single-particle excitation spectra and the excitonic density of states, which would be instrumental for interpretation of the coherent light emission measurements in the excitonic system. 

\appendix

\section{\label{sec:Section_A} Effective actions }
%
\subsection{\label{sec:Section_A_1} Fermionic action}
%
We would like now to derive the effective action for fermions.
Our starting point is the partition function given in the Eq.(\ref{Equation_28}) derived with the help of the U$(1)$ gauge transformation as it is introduced in the Section \ref{sec:Section_3} and which can be written as
\begin{eqnarray}
{\cal{Z}}=\int\left[{\cal{D}}\bar{a}{\cal{D}}a\right]\left[{\cal{D}}\bar{b}{\cal{D}}b\right] e^{-{\cal{S}}_{\rm eff}[{\bar{a},a,\bar{b},b}]},
\label{Equation_A1}
\end{eqnarray}
where the effective fermionic action ${\cal{S}}_{\rm eff}[{\bar{a},a,\bar{b},b}]$ in the exponential is defined as 
\begin{eqnarray}
{\cal{S}}_{\rm eff}[{\bar{a},a,\bar{b},b}]=-\ln\int\left[{\cal{D}}\varphi\right]e^{-{\cal{S}}[\bar{a},a,{\bar{b}},b,\varphi]}.
\label{Equation_A2}
\end{eqnarray}
Furthermore, we expand the logarithm keeping only the terms up to second order in ${\cal{S}}$. As a result we obtain
\begin{eqnarray}
{\cal{S}}_{\rm eff}[{\bar{a},a,\bar{b},b}]=&&{\cal{S}}_{0}+\left\langle{\cal{S}}\right\rangle_{{\cal{S}}_{\rm eff}[\varphi]}
\nonumber\\
&&
-\frac{1}{2}\left[\left\langle{\cal{S}}^{2}\right\rangle_{{\cal{S}}_{\rm eff}[\varphi]}-\left\langle{\cal{S}}\right\rangle^{2}_{{\cal{S}}_{\rm eff}[\varphi]}\right].
\label{Equation_A3}
\nonumber\\
\ \ \ 
\end{eqnarray}
Here, the averages with respect to the phase variables are defined as
\begin{eqnarray}
\left\langle...\right\rangle_{{\cal{S}}_{\rm eff}[\varphi]}=\frac{\int\left[{\cal{D}}\varphi\right]...e^{-{\cal{S}}_{\rm eff}[\varphi]}}{\int\left[{\cal{D}}\varphi\right]e^{-{\cal{S}}_{\rm eff}[\varphi]}}.
\label{Equation_A4}
\end{eqnarray}
%
\subsection{\label{sec:Section_A_2} Phase action}
%
In a similar way, the integration over the fermions in Eq.(\ref{Equation_28}) gives the effective action for the phase sector. The partition function in this case is
\begin{eqnarray}
{\cal{Z}}=\int\left[{\cal{D}}\varphi\right]e^{-{\cal{S}}_{\rm eff}[\varphi]},
\label{Equation_A5}
\end{eqnarray}
where the effective phase action in the exponential is 
\begin{eqnarray}
{\cal{S}}_{\rm eff}[\varphi]=-\ln\int\left[{\cal{D}}\bar{a}{\cal{D}}a\right]\left[{\cal{D}}\bar{b}{\cal{D}}b\right] e^{-{\cal{S}}}.
\label{Equation_A6}
\end{eqnarray}
Again, by expanding the logarithm in the Eq.(\ref{Equation_A5}), we will have up to second order in ${\cal{S}}$
\begin{eqnarray}
{\cal{S}}_{\rm eff}[\varphi]
=&&\tilde{\cal{S}}_{0}+\left\langle{\cal{S}}\right\rangle_{{\cal{S}}_{\rm eff}\left[{\bar{a},a,\bar{b},b}\right]}
\nonumber\\
&&-\frac{1}{2}\left[\left\langle{\cal{S}}^{2}\right\rangle_{{\cal{S}}_{\rm eff}{\left[{\bar{a},a,\bar{b},b}\right]}}-\left\langle{\cal{S}}\right\rangle^{2}_{{\cal{S}}_{\rm eff}\left[{\bar{a},a,\bar{b},b}\right]}\right],
\label{Equation_A7}
\nonumber\\
\ \ \ 
\end{eqnarray}
where $\tilde{\cal{S}}_{0}$ is an unimportant constant. 
Here the fermionic average $\langle ... \rangle_{{\cal{S}}_{\rm eff}}$ is given by
\begin{eqnarray}
\left\langle...\right\rangle_{{\cal{S}}_{\rm eff}\left[{\bar{a},a,\bar{b},b}\right]}=\frac{\int\left[{\cal{D}}\bar{a}{\cal{D}}a\right]\left[{\cal{D}}\bar{b}{\cal{D}}b\right]...e^{-{\cal{S}}_{\rm eff}\left[{\bar{a},a,\bar{b},b}\right]}}{\int\left[{\cal{D}}\bar{a}{\cal{D}}a\right]\left[{\cal{D}}\bar{b}{\cal{D}}b\right]e^{-{\cal{S}}_{\rm eff}\left[{\bar{a},a,\bar{b},b}\right]}}.
\nonumber\\
\label{Equation_A8}
\end{eqnarray}

The Eq.(\ref{Equation_A3}) is important for deriving the excitonic phase-stiffness parameter. 
We present here derivation of the terms in the effective phase action, which are proportional to $t\tilde{t}$, in  Eq.(\ref{Equation_A6}). The derivation of the other term, proportional to $\tilde{t}t$, is very similar. Using Eq.(\ref{Equation_A6})  and Eq.(\ref{Equation_26}) we get for the mixed term $t\tilde{t}$
\begin{eqnarray}
&&-\frac{1}{2}\sum_{\left\langle{\bf{r}}_{1},{\bf{r}}'_{1}\right\rangle}\sum_{\left\langle{\bf{r}}_{2},{\bf{r}}'_{2}\right\rangle}\int^{\beta}_{0}d\tau d\tau' \left[ t({\bf{r}}_{1}{\bf{r}}'_{1})\tilde{t}({\bf{r}}_{2}{\bf{r}}'_{2})\times
\right.
\nonumber\\
\nonumber\\
&&\left.\left\langle \bar{a}({\bf{r}}_{1}\tau){a}({\bf{r}}'_{1}\tau)\bar{b}({\bf{r}}_{2}\tau'){b}({\bf{r}}'_{2}\tau')\right\rangle\times \right.
\nonumber\\
\nonumber\\
&&\left.
\times e^{-i\left[\varphi({\bf{r}}_{1}\tau)-\varphi({\bf{r}}'_{1}\tau)\right]}e^{-i\left[\varphi({\bf{r}}_{2}\tau')-\varphi({\bf{r}}'_{2}\tau')\right]}\right.
\nonumber\\
\nonumber\\
&&\left.+t({\bf{r}}_{1}{\bf{r}}'_{1})\tilde{t}({\bf{r}}'_{2}{\bf{r}}_{2})\left\langle \bar{a}({\bf{r}}_{1}\tau){a}({\bf{r}}'_{1}\tau)\bar{b}({\bf{r}}'_{2}\tau'){b}({\bf{r}}_{2}\tau')\right\rangle\times \right.
\nonumber\\
\nonumber\\
&&\left.
\times e^{-i\left[\varphi({\bf{r}}_{1}\tau)-\varphi({\bf{r}}'_{1}\tau)\right]}e^{i\left[\varphi({\bf{r}}_{2}\tau')-\varphi({\bf{r}}'_{2}\tau')\right]}\right.
\nonumber\\
\nonumber\\
&&\left.+t({\bf{r}}'_{1}{\bf{r}}_{1})\tilde{t}({\bf{r}}_{2}{\bf{r}}'_{2})\left\langle \bar{a}({\bf{r}}'_{1}\tau){a}({\bf{r}}_{1}\tau)\bar{b}({\bf{r}}_{2}\tau'){b}({\bf{r}}'_{2}\tau')\right\rangle\times \right.
\nonumber\\
\nonumber\\
&&\left.
\times e^{i\left[\varphi({\bf{r}}_{1}\tau)-\varphi({\bf{r}}'_{1}\tau)\right]}e^{-i\left[\varphi({\bf{r}}_{2}\tau')-\varphi({\bf{r}}'_{2}\tau')\right]}\right.
\nonumber\\
\nonumber\\
&&\left.+t({\bf{r}}'_{1}{\bf{r}}_{1})\tilde{t}({\bf{r}}'_{2}{\bf{r}}_{2})\left\langle \bar{a}({\bf{r}}'_{1}\tau){a}({\bf{r}}_{1}\tau)\bar{b}({\bf{r}}'_{2}\tau'){b}({\bf{r}}_{2}\tau')\right\rangle\times \right.
\nonumber\\
\nonumber\\
&&\left.
\times e^{i\left[\varphi({\bf{r}}_{1}\tau)-\varphi({\bf{r}}'_{1}\tau)\right]}e^{i\left[\varphi({\bf{r}}_{2}\tau')-\varphi({\bf{r}}'_{2}\tau')\right]}\right].
\label{Equation_A9}
\end{eqnarray}
As an example, we give the Wick averaging result of the first four-fermion term in the expression of Eq.(\ref{Equation_A9})
\begin{eqnarray}
&&\left\langle \bar{a}({\bf{r}}_{1}\tau){a}({\bf{r}}'_{1}\tau)\bar{b}({\bf{r}}_{2}\tau'){b}({\bf{r}}'_{2}\tau')\right\rangle=
\nonumber\\
&&
=\left\langle \bar{a}({\bf{r}}_{1}\tau){a}({\bf{r}}'_{1}\tau)\right\rangle \left\langle\bar{b}({\bf{r}}_{2}\tau'){b}({\bf{r}}'_{2}\tau')\right\rangle
\nonumber\\
&&-\left\langle\bar{a}({\bf{r}}_{1}\tau)\bar{b}({\bf{r}}_{2}\tau')\right\rangle \left\langle a({{\bf{r}}'_{1}{\tau}})b({\bf{r}}'_{2}\tau')\right\rangle
\nonumber\\
&&+\left\langle\bar{a}({\bf{r}}_{1}\tau){b}({\bf{r}}'_{2}\tau')\right\rangle \left\langle a({{\bf{r}}'_{1}{\tau}})\bar{b}({\bf{r}}_{2}\tau')\right\rangle=
\nonumber\\
&&={\cal{G}}^{aa}({\bf{r}}'_{1}-{\bf{r}}_{1},0){\cal{G}}^{bb}({\bf{r}}'_{2}-{\bf{r}}_{2},0)
\nonumber\\
&&-{\cal{G}}^{ab}({\bf{r}}_{1}-{\bf{r}}'_{2},\tau-\tau'){\cal{G}}^{ba}({\bf{r}}_{2}-{\bf{r}}'_{1},\tau'-\tau).
\label{Equation_A10}
\end{eqnarray}
We kept in Eq.(\ref{Equation_A9}) only the terms proportional to excitonic gap. We neglected other terms like $\left\langle\bar{a}({\bf{r}}\tau)\bar{b}({\bf{r}}'\tau')\right\rangle$ or $\left\langle{a}({\bf{r}}\tau){b}({\bf{r}}'\tau')\right\rangle$, which vanish due to the symmetry  of the action in Eq.(\ref{Equation_62}). Contributions, proportional to fermionic densities $\left\langle\bar{a}({\bf{r}}\tau)a({\bf{r}}'\tau')\right\rangle$ and $\left\langle\bar{b}({\bf{r}}\tau)b({\bf{r}}'\tau')\right\rangle$ could be also omitted, since they are not contributing directly to the excitonic pair formation.

After calculating all averages in Eq.(\ref{Equation_A9}) and recombining them with the similar terms coming from the component proportional to $\tilde{t}t$, we obtain the relevant portion of the phase action in the form
\begin{widetext}
\begin{eqnarray}
{\cal{S}}_{J}\left[\varphi\right]=-&&-\int^{\beta}_{0}d\tau \int^{\beta}_{0}d\tau' \sum_{{\bf{r}},{\bf{r}}'}\left\{J({\bf{r}}\tau,{\bf{r}}'\tau')\cos\left[\varphi({\bf{r}}\tau)+\varphi({\bf{r}}\tau')-\varphi({\bf{r}}'\tau)-\varphi({\bf{r}}'\tau')\right]\right.
\nonumber\\
&&\left.+{\cal{G}}_{ab}({\bf{0}},\tau-\tau'){\cal{G}}_{ba}({\bf{0}},\tau'-\tau)\cos\left[\varphi({\bf{r}}\tau)-\varphi({\bf{r}}\tau')-\varphi({\bf{r}}'\tau)+\varphi({\bf{r}}'\tau')\right]\right\},
\label{Equation_A11}
\end{eqnarray}
\end{widetext}
which contains the phase-stiffness parameter $J$
\begin{eqnarray}
J({\bf{r}}\tau,{\bf{r}}'\tau')={4t\tilde{t}}{\cal{G}}_{ab}({\bf{r}}-{\bf{r}}',\tau-\tau'){\cal{G}}_{ba}({\bf{r}}-{\bf{r}}',\tau'-\tau).
\nonumber\\
\ \ \ 
\label{Equation_A12}
\end{eqnarray}

Furthermore, in order to simplify the non-local (in time variables) effective phase action in Eq.(\ref{Equation_A11}), we resort to the gradient expansion of the phase field in the form 
\begin{eqnarray}
\varphi({\bf{r}}\tau')=\varphi({\bf{r}}\tau)+\left(\tau'-\tau\right)\partial_{\tau}\varphi({\bf{r}}\tau)+O\left[\left(\tau'-\tau\right)^{2}\right].
\label{Equation_A13}
\nonumber\\
\end{eqnarray}
As a result we can deduce the phase-stiffness parameter $J$ in the form 
\begin{eqnarray}
J={4t\tilde{t}}\int^{\beta}_{0}d\tau'{\cal{G}}_{ab}({\bf{r}}-{\bf{r}}',\tau-\tau'){\cal{G}}_{ba}({\bf{r}}-{\bf{r}}',\tau'-\tau),
\nonumber\\
\ \ \ 
\label{Equation_A14}
\end{eqnarray}
while the phase  action in E.(\ref{Equation_A11}) simplifies to that given in Eq.(\ref{Equation_61}), which now is local in time variable $\tau$.
For the product of the anomalous propagators in Eq.(\ref{Equation_A14}) we have 
\begin{widetext}
\begin{eqnarray}
&&{\cal{G}}_{ab}(\tau-\tau'){\cal{G}}_{ba}(\tau'-\tau)
=\frac{4\Delta^{2}U^{2}}{z^{2}(\beta{N})^{2}}\sum_{\substack{{\bf{k}},{\bf{k}}'}}
\sum_{\nu_{n}, \nu'_{n}}\frac{\epsilon\left({{\bf{k}}}\right)\epsilon\left({{\bf{k}}}'\right)e^{-i(\nu'_{n}-\nu_{n})\delta}}{\left[{\cal{E}}^{a}_{{\bf{k}}}(\nu_{n}){\cal{E}}^{b}_{{\bf{k}}}(\nu_{n})-|\Delta|^{2}\right]\left[{\cal{E}}^{a}_{{\bf{k}}'}(\nu_{n}'){\cal{E}}^{b}_{{\bf{k}}'}(\nu_{n}')-|\Delta|^{2}\right]}.
\nonumber\\
\ \ \ \ 
\label{Equation_A15}
\end{eqnarray}
\end{widetext}
Here $z=6$ is the number of the n.n. sites on the 3D cubic lattice.
After integrating over the imaginary time $\tau'$ in Eq.(\ref{Equation_A14}), we perform the Matsubara frequency summations in Eq.(\ref{Equation_A15}) and obtain the phase-stiffness parameter $J$ in Eq.(\ref{Equation_A14}) in the final form
\begin{widetext}
\begin{eqnarray}
J=\frac{16\Delta^{2}t\tilde{t}}{z^{2}{N^{2}}}\sum_{{\bf{k}},{\bf{k}}'}\frac{\epsilon\left({{\bf{k}}}\right)\epsilon\left({{\bf{k}}}'\right)}{{\sqrt{\xi^{2}_{{\bf{k}}}+4\Delta^{2}}}}\cdot\left[\Lambda_{1}({\bf{k}},{\bf{k}}')\tanh\left(\frac{\beta E^{+}_{{\bf{k}}}}{2}\right)-\Lambda_{2}({\bf{k}},{\bf{k}}')\tanh\left(\frac{\beta E^{-}_{{\bf{k}}}}{2}\right)\right].
\label{Equation_A16}
\end{eqnarray}
\end{widetext}
The parameters $\Lambda_{1}({\bf{k}},{\bf{k}}')$ and $\Lambda_{2}({\bf{k}},{\bf{k}}')$ entering in Eq.(\ref{Equation_A16}) are given by
\begin{eqnarray}
\Lambda_{1}({\bf{k}},{\bf{k}}')=\frac{1}{E^{+}({\bf{k}}) - E^{+}({\bf{k}}')}\cdot\frac{1}{E^{+}({\bf{k}}) - E^{-}({\bf{k}}')},
\nonumber\\
\Lambda_{2}({\bf{k}},{\bf{k}}')=\frac{1}{E^{-}({\bf{k}}) - E^{-}({\bf{k}}')}\cdot\frac{1}{E^{-}({\bf{k}}) - E^{+}({\bf{k}}')}.
\label{Equation_A17}
\nonumber\\
\ \ \ 
\end{eqnarray}
The summations over the ${\bf{k}}$ wave vectors in Eq.(\ref{Equation_A16}) could be transformed into the integrations with the help of the density of states given in Eq.(\ref{Equation_59}) (see the Section \ref{sec:Section_4_2}).
As we see, Eq.(\ref{Equation_A16}) relates the parameter $J$ with the local pairing gap $\Delta$. The numerical evaluations of the expression in Eq.(\ref{Equation_A16}) for $T=0$ K are presented in Fig.~\ref{fig:Fig_10} and discussed in the Section \ref{sec:Section_5} of the present paper.
%
\section{\label{Section_B} The action ${\cal{S}}_{\lambda}\left[\bar{z},z\right]$}
%
The action in Eq.(\ref{Equation_70}) is quartic in unimodular $z$-field and could be decoupled with the help of the MF-like decoupling procedure
\begin{eqnarray}
\left[\bar{z}\left({\bf{r}}\tau\right)z\left({\bf{r}}'\tau\right)+c.c.\right]^{2}\rightarrow 4\left\langle{\bar{z}\left({\bf{r}}\tau\right)z\left({\bf{r}}'\tau\right)}\right\rangle\left[\bar{z}\left({\bf{r}}\tau\right)z\left({\bf{r}}'\tau\right)+c.c.\right]
\label{Equation_B1}
\end{eqnarray}
Then we get
\begin{eqnarray}
{\cal{S}}_{J}[\bar{z},z]=-2g_{B}J\int^{\beta}_{0}d\tau \sum_{\left\langle{\bf{r}},{\bf{r}}'\right\rangle}\bar{z}({\bf{r}}\tau)z({\bf{r}}'\tau).
\label{Equation_B2}
\end{eqnarray}
Now we will derive the action given in Eq.(\ref{Equation_74}). We start with the partition function given in Eq.(\ref{Equation_71}). We introduce the Fadeev-Popov resolution for the delta functions in Eq.(\ref{Equation_71}) by introducing the ghost-fields ${\eta}({\bf{r}}\tau)$ and ${\bar{\eta}}({\bf{r}}\tau)$ as
\begin{eqnarray}
\delta(z-e^{i\varphi\left({\bf{r}}\tau\right)})=\int\left[{\cal{D}}\bar{\eta}\right]e^{i\int^{\beta}_{0}d\tau\sum_{{\bf{r}}}\bar{\eta}({\bf{r}}\tau)\left(z-e^{i\varphi\left({\bf{r}}\tau\right)}\right)},
\nonumber\\
\delta(\bar{z}-e^{-i\varphi\left({\bf{r}}\tau\right)})=\int\left[{\cal{D}}{\eta}\right]e^{i\int^{\beta}_{0}d\tau\sum_{{\bf{r}}}{\eta}({\bf{r}}\tau)\left(\bar{z}-e^{-i\varphi\left({\bf{r}}\tau\right)}\right)}.
\label{Equation_B3}
\end{eqnarray}
Then we write
\begin{eqnarray}
&&e^{i\int^{\beta}_{0}d\tau\sum_{{\bf{r}}}\bar{\eta}({\bf{r}}\tau)\left(z-e^{i\varphi\left({\bf{r}}\tau\right)}\right)}=\lim_{N\rightarrow \infty}\prod^{N}_{n=1}\prod_{{\bf{r}}}e^{i\epsilon_{n}\bar{\eta}({\bf{r}}\tau)\left(z-e^{i\varphi\left({\bf{r}}\tau\right)}\right)}=
\nonumber\\
&&=\lim_{N\rightarrow \infty}\prod^{N}_{n=1}\prod_{{\bf{r}}}e^{i\epsilon_{n}\bar{\eta}({\bf{r}}\tau)z\left({\bf{r}}\tau\right)}\left[1-i\epsilon_{n}\bar{\eta}({\bf{r}}\tau)e^{i\varphi({\bf{r}}\tau)}+\frac{1}{2!}\left(-i\epsilon_{n}\bar{\eta}({\bf{r}}\tau)e^{i\varphi({\bf{r}}\tau)}\right)^{2}+\frac{1}{3!}\left(-i\epsilon_{n}\bar{\eta}({\bf{r}}\tau)e^{i\varphi({\bf{r}}\tau)}\right)^{3}+...\right].
\nonumber\\
\label{Equation_B4}
\end{eqnarray}
We can write also an analogue expression for the exponential $e^{i\int^{\beta}_{0}d\tau\sum_{{\bf{r}}}{\eta}\left(\bar{z}-e^{-i\varphi\left({\bf{r}}\tau\right)}\right)}$. Thereby, we have
\begin{eqnarray}
&&e^{i\int^{\beta}_{0}d\tau\sum_{{\bf{r}}}\bar{\eta}\left(z-e^{i\varphi\left({\bf{r}}\tau\right)}\right)}e^{i\int^{\beta}_{0}d\tau\sum_{{\bf{r}}}{\eta}({\bf{r}}\tau)\left(\bar{z}-e^{-i\varphi\left({\bf{r}}\tau\right)}\right)}=\lim_{N\rightarrow \infty}\prod^{N}_{n,m=1}\prod_{{\bf{r}},{\bf{r}}'}\left[1+\left(-i\epsilon_{n}\bar{\eta}e^{i\varphi({\bf{r}}\tau)}\right)\left(-i\epsilon_{m}{\eta}({\bf{r}}\tau)e^{-i\varphi({\bf{r}}'\tau')}\right)\right.
\nonumber\\
&&\left.+\left(\frac{1}{2!}\right)^{2}\left(-i\epsilon_{n}\bar{\eta}({\bf{r}}\tau)e^{i\varphi({\bf{r}}\tau)}\right)^{2}\left(-i\epsilon_{m}{\eta}({\bf{r}}\tau)e^{-i\varphi({\bf{r}}'\tau')}\right)^{2}+...\right]e^{i\int^{\beta}_{0}d\tau\sum_{{\bf{r}}}\bar{\eta}({\bf{r}}\tau)z({\bf{r}}\tau)}e^{i\int^{\beta}_{0}d\tau\sum_{{\bf{r}}}z({\bf{r}}\tau){\eta}({\bf{r}}\tau)}.
\label{Equation_B5}
\end{eqnarray}
We put now the expression in Eq.(\ref{Equation_B5}) into the partition function in Eq.(\ref{Equation_71}) and we integrate out the phase variables $\varphi\left({\bf{r}}\tau\right)$
\begin{eqnarray}
{\cal{Z}}=\lim_{N\rightarrow \infty}\prod^{N}_{n,m=1}\prod_{{\bf{r}},{\bf{r}}'}\int\left[{\cal{D}}\lambda\right]\left[{\cal{D}}\bar{z}{\cal{D}}z\right]\left[{\cal{D}}\bar{\eta}{\cal{D}}\eta\right]e^{2g_{B}J\int^{\beta}_{0}d\tau \sum_{\left\langle{\bf{r}},{\bf{r}}'\right\rangle}\bar{z}({\bf{r}}\tau)z({\bf{r}}'\tau)}e^{i\int^{\beta}_{0}d\tau\sum_{{\bf{r}}}\lambda\left(|z({\bf{r}}\tau)|^{2}-1\right)}\times
\nonumber\\
\times\left[1-\frac{1}{1!}{\bar{\eta}}({\bf{r}}\tau)\eta({\bf{r}}'\tau')\epsilon_{n}\epsilon_{m}\frac{\left\langle e^{i\left[\varphi({\bf{r}}\tau)-\varphi({\bf{r}}'\tau')\right]}\right\rangle}{1!}+\frac{1}{2!}{\bar{\eta}}^{2}({\bf{r}}\tau)\eta^{2}({\bf{r}}'\tau')\epsilon^{2}_{n}\epsilon^{2}_{m}\frac{\left\langle e^{i2\left[\varphi({\bf{r}}\tau)-\varphi({\bf{r}}'\tau')\right]}\right\rangle}{2!}-\right.
\nonumber\\
\left.-\frac{1}{3!}{\bar{\eta}}^{3}({\bf{r}}\tau)\eta^{3}({\bf{r}}'\tau')\epsilon^{3}_{n}\epsilon^{3}_{m}\frac{\left\langle e^{i3\left[\varphi({\bf{r}}\tau)-\varphi({\bf{r}}'\tau')\right]}\right\rangle}{3!}+...\right].
\label{Equation_B6}
\end{eqnarray}
The phase averages in Eq.(\ref{Equation_B6}) are given as 
\begin{eqnarray}
\left\langle e^{in\left[\varphi({\bf{r}}\tau)-\varphi({\bf{r}}'\tau')\right]}\right\rangle=\frac{\int\left[{\cal{D}}\varphi\right]e^{-{\cal{S}}_{0}\left[\varphi\right]}e^{in\left[\varphi({\bf{r}}\tau)-\varphi({\bf{r}}'\tau')\right]}}{\int\left[{\cal{D}}\varphi\right]e^{-{\cal{S}}_{0}\left[\varphi\right]}}.
\label{Equation_B7}
\end{eqnarray}
On the other can decouple the expression $\left\langle e^{in\left[\varphi({\bf{r}}\tau)-\varphi({\bf{r}}'\tau')\right]}\right\rangle$ using the MF like cumulant averaging procedure and we obtain $\left\langle e^{in\left[\varphi({\bf{r}}\tau)-\varphi({\bf{r}}'\tau')\right]}\right\rangle=\left\langle e^{i\left[\varphi({\bf{r}}\tau)-\varphi({\bf{r}}'\tau')\right]}\right\rangle n!$.
Then, we rewrite the expression in Eq.(\ref{Equation_B6}) in the more simple form
\begin{eqnarray}
{\cal{Z}}=\int\left[{\cal{D}}\lambda\right]\left[{\cal{D}}\bar{z}{\cal{D}}z\right]\left[{\cal{D}}\bar{\eta}{\cal{D}}\eta\right]e^{2g_{B}J\int^{\beta}_{0}d\tau \sum_{\left\langle{\bf{r}},{\bf{r}}'\right\rangle}\bar{z}({\bf{r}}\tau)z({\bf{r}}'\tau)}e^{i\int^{\beta}_{0}d\tau\sum_{{\bf{r}}}\lambda\left(|z({\bf{r}}\tau)|^{2}-1\right)}\times
\nonumber\\
\times e^{-\sum_{{\bf{r}},{\bf{r}}'}\int^{\beta}_{0}d\tau\int^{\beta}_{0}d\tau'\bar{\eta}\left({\bf{r}}\tau\right)\gamma\left({\bf{r}}\tau,{\bf{r}}'\tau'\right)\eta\left({\bf{r}}'\tau'\right)+i\int^{\beta}_{0}d\tau\sum_{{\bf{r}}}\bar{\eta}({\bf{r}}\tau)z({\bf{r}}\tau)+i\int^{\beta}_{0}d\tau\sum_{{\bf{r}}}\bar{z}({\bf{r}}\tau){\eta}({\bf{r}}\tau)},
\label{Equation_B8}
\end{eqnarray}
where we introduced the phase-correlation function $\gamma\left({\bf{r}}\tau,{\bf{r}}'\tau'\right)=\left\langle e^{i\left[\varphi({\bf{r}}\tau)-\varphi({\bf{r}}'\tau')\right]}\right\rangle$. Now, we integrate out the bosonic $\eta$-field by employing the HS complex transformation for bosons
\begin{eqnarray}
&&\int{\frac{1}{N}\prod_{i}d\bar{\zeta}_{i}d\zeta{i}}e^{-\sum_{ij}\bar{\zeta}_{i}A^{-1}_{ij}\zeta_{j}+\sum_{i}[\bar{z}_{i}{\zeta}_{i}+z_{i}\bar{\zeta}_{i}]}
\nonumber\\
&&=\left[\det{A}^{-1}\right]^{-1}e^{\sum_{ij}\bar{z}_{i}A_{ij}z_{j}},
\label{Equation_B9}
\end{eqnarray}
we get
\begin{eqnarray}
\int\left[{\cal{D}}\bar{\eta}{\cal{D}}\eta\right]e^{-\sum_{{\bf{r}},{\bf{r}}'}\int^{\beta}_{0}d\tau\int^{\beta}_{0}d\tau'\bar{\eta}\left({\bf{r}}\tau\right)\gamma\left({\bf{r}}\tau,{\bf{r}}'\tau'\right)\eta\left({\bf{r}}'\tau'\right)+i\int^{\beta}_{0}d\tau\sum_{{\bf{r}}}\bar{\eta}({\bf{r}}\tau)z({\bf{r}}\tau)+i\int^{\beta}_{0}d\tau\sum_{{\bf{r}}}\bar{z}({\bf{r}}\tau){\eta}({\bf{r}}\tau)}\approx
\nonumber\\
\approx e^{-\sum_{{\bf{r}},{\bf{r}}'}\int^{\beta}_{0}d\tau\int^{\beta}_{0}d\tau'\bar{z}\left({\bf{r}}\tau\right)\gamma^{-1}\left({\bf{r}}\tau,{\bf{r}}'\tau'\right)z\left({\bf{r}}'\tau'\right)}.
\label{Equation_B10}
\end{eqnarray}
For the partition function in Eq.(\ref{Equation_B8}) we have
\begin{eqnarray}
{\cal{Z}}=\int\left[{\cal{D}}\lambda\right]\left[{\cal{D}}\bar{z}{\cal{D}}z\right]e^{2g_{B}J\int^{\beta}_{0}d\tau \sum_{\left\langle{\bf{r}},{\bf{r}}'\right\rangle}\bar{z}({\bf{r}}\tau)z({\bf{r}}'\tau)}e^{i\int^{\beta}_{0}d\tau\sum_{{\bf{r}}}\lambda\left(|z({\bf{r}}\tau)|^{2}-1\right)}\times
\nonumber\\
\times e^{-\sum_{{\bf{r}},{\bf{r}}'}\int^{\beta}_{0}d\tau\int^{\beta}_{0}d\tau'\bar{z}\left({\bf{r}}\tau\right)\gamma^{-1}\left({\bf{r}}\tau,{\bf{r}}'\tau'\right)z\left({\bf{r}}'\tau'\right)}
\label{Equation_B11}
\end{eqnarray}
or, similarly,
\begin{eqnarray}
{\cal{Z}}=\int\left[{\cal{D}}\lambda\right]\left[{\cal{D}}\bar{z}{\cal{D}}z\right]e^{-\sum_{{{\bf{r}}},{\bf{r}}'}\int^{\beta}_{0}d\tau\int^{\beta}_{0}d\tau'\bar{z}({\bf{r}}\tau){\cal{G}}^{-1}_{z}({\bf{r}}\tau,{\bf{r}}'\tau')z({\bf{r}}'\tau')},
\label{Equation_B12}
\end{eqnarray}
where ${\cal{G}}^{-1}_{z}({\bf{r}}\tau,{\bf{r}}'\tau')$ is the inverse of the real-space bosonic Green-function matrix. 
\begin{eqnarray}
{\cal{G}}^{-1}_{z}({\bf{r}}\tau,{\bf{r}}'\tau')=-2g_{B}J\delta(\tau-\tau')\delta({\bf{r}}-{\bf{r}}'-{\bf{d}})+\lambda\delta\left({{\bf{r}}-{\bf{r}}'}\right)\delta(\tau-\tau')+\gamma^{-1}({\bf{r}}\tau,{\bf{r}}'\tau').
\label{Equation_B13}
\end{eqnarray}
In fact, the phase correlation function $\gamma\left({\bf{r}}\tau,{\bf{r}}'\tau'\right)$ has the form
\begin{eqnarray}
\gamma\left({\bf{r}}\tau,{\bf{r}}'\tau'\right)=\delta\left({{\bf{r}}-{\bf{r}}'}\right)e^{-\frac{U}{\beta}\sum^{\infty}_{n=1}\frac{1-\cos\left[\omega_{n}\left(\tau-\tau'\right)\right]}{\omega^{2}_{n}}}\times
\nonumber\\
\times \sum_{\left\{m\right\}}e^{-\frac{U\beta}{4}\left[m({\bf{r}})-\frac{2\bar{\mu}}{U}\right]^{2}-\frac{U}{2}\left(m-\frac{2\bar{\mu}}{U}\right)\left(\tau-\tau'\right)},
\label{Equation_B14}
\end{eqnarray}
where $\left\{m\right\}$ forms an infinite set of U(1) winding numbers (see the Section \ref{sec:Section_3_1}). Transforming the $z$-variables into the Fourier space (see the Section \ref{sec:Section_5}) we can write the partition function in Eq.(\ref{Equation_B12}) as
\begin{eqnarray}
{\cal{Z}}=\int\left[{\cal{D}}\lambda\right]\left[{\cal{D}}\bar{z}{\cal{D}}z\right]e^{-\frac{1}{\beta{N}}\sum_{{\bf{k}},\omega_{n}}{\bar{z}}({\bf{k}}\omega_{n}){\cal{G}}^{-1}_{z}({\bf{k}}\omega_{n})z({\bf{k}}\omega_{n})}
\label{Equation_B15}
\end{eqnarray}
and now ${\cal{G}}^{-1}_{z}({\bf{k}}\omega_{n})$ is
\begin{eqnarray}
{\cal{G}}^{-1}_{z}({\bf{k}}\omega_{n})=\gamma^{-1}(\omega_{n})-4g_{B}J-\lambda,
\label{Equation_B16}
\end{eqnarray}
where $\gamma^{-1}(\omega_{n})$ is the inverse of the Fourier transformation $\gamma(\omega_{n})$ of $\gamma(\tau-\tau')$ given in Eq.(\ref{Equation_77}) in the Section \ref{sec:Section_5}.
%

%

\begin{references}
%
\bibitem{Moskalenko} Moskalenko S. A., Snoke D. W., \textit{Bose-Einstein Condensation of Excitons and Biexcitons} (Cambridge University Press, 2005).
\bibitem{Keldysh_1} L. V. Keldysh and A. N. Kozlov, Zh. Eksp. Teor. Fiz. \textbf{54}, 978–993 (1968), Sov. Phys. JETP \textbf{27}, 521–528 (1968).
\bibitem{Neuenschwander} J. Neuenschwander, P. Wachter, Phys. Rev. B \textbf{41} (1990) 12693.
\bibitem{Wachter_1} B. Bucher, P. Steiner, and P. Wachter, Phys. Rev. Lett. \textbf{67}, 2717 (1991).
\bibitem{Wachter_2} P. Wachter, Solid State Commun. \textbf{118}, 645, (2001).
\bibitem{Wachter} P. Wachter, B. Bucher, J. Malar, Phys. Rev. B \textbf{69} (2004) 094502.
\bibitem{Keldysh_2} L. V. Keldysh and Y. V. Kopaev, Fiz. Tverd. Tela (Leningrad) \textbf{6},
2791 (1964) [Sov. Phys. Solid State \textbf{6}, 2219 (1965)].
\bibitem{Cloizeaux} J. des Cloizeaux, J. Chem. Phys. Solids. 26, 259 (1965).
\bibitem{Kohn} W. Kohn, in Many Body Physics, edited by C. de Witt and R. Balian (Gordon \&
 Breach, New York, 1968).
\bibitem{Jerome}D. J\'{e}rome, T. M. Rice, and W. Kohn, Phys. Rev. \textbf{158}, 462 (1967).
\bibitem{Keldysh_3} A. Griffin, D.W. Snoke and S. Stringari, eds \textit{Bose-Einstein Condensation}, (Cambridge University Press, 1995).
\bibitem{Halperin} B. I. Halperin and T. M. Rice, in Solid State Physics, edited by F. Seitz, D. Turnbull, and H. Ehrenreich, Academic, New York, (1967).
(Cambridge U. Press, Cambridge, 1995).
\bibitem{Wakisaka} Y. Wakisaka, T. Sudayama, K. Takubo, T. Mizokawa, M. Arita, H. Namatame, M. Taniguchi, N. Katayama, M. Nohara, and H. Takagi, Phys. Rev. Lett. \textbf{103}, 026402 (2009).
\bibitem{Berger}H. Cercellier, C. Monney, F. Clerc, C. Battaglia, L. Despont, M. G. Garnier, H. Beck, P. Aebi, L. Patthey, H. Berger and L. Forr\'{o}, Phys. Rev. Lett. \textbf{99}, 146403 (2007).

\bibitem{Kaneko} T. Kaneko, T. Toriyama, T. Konishi and Y. Ohta, Phys. Rev. B \textbf{87}, 035121 (2013).
\bibitem{Bardeen} J. Bardeen, L. N. Cooper, and J. R. Schrieffer, Phys. Rev. \textbf{106}, 162, (1957).
\bibitem{Ihle} D. Ihle, M. Pfafferott, E. Burovski, F. X. Bronold, and H. Fehske, Phys. Rev. B \textbf{78}, 193103 (2008).
\bibitem{Pethick} C. J. Pethick and H. Smith, \textit{Bose Einstein Condensation in Dilute Gases}, (Cambridge University Press, Cambridge, 2001).
\bibitem{Chen} Qijin Chen, Jelena Stajic, Shina Tan, Kathryn Levin, Physics Reports \textbf{412}, 1-88 (2005).
\bibitem{Kremp} D. Kremp, D. Semkat, and K. Henneberger, Phys. Rev. B \textbf{78}, 125315 2008.
\bibitem{Bronold} F. X. Bronold and H. Fehske, Phys. Rev. B \textbf{74}, 165107 (2006).  
\bibitem{Zenker_1} B. Zenker, D. Ihle, F. X. Bronold, and H. Fehske, Phys. Rev. B \textbf{81}, 115122 March (2010). 
\bibitem{Zenker_2}B. Zenker, D. Ihle, F. X. Bronold, and H. Fehske,  Phys. Rev. B \textbf{83}, 235123 (2011).
\bibitem{Seki} K. Seki, R. Eder, and Y. Ohta,  { Phys. Rev. B}, \textbf{84}, 245106 (2011).
\bibitem{Golosov} D. I. Golosov, Phys. Rev. B, \textbf{86}, 155134 (2012).
\bibitem{Farkasovsky_1} P. Farka\v{s}ovsk\'{y}, Phys. Rev. B \textbf{77}, 155130 (2008).
\bibitem{Zenker_3} B. Zenker, D. Ihle, F. X. Bronold, and H. Fehske, Phys. Rev. B \textbf{85}, 121102 (2012).
\bibitem{Batista_1} C. D. Batista, Phys. Rev. Lett. \textbf{89}, 166403, (2002).
\bibitem{Batista_2} C. D. Batista, J. E. Gubernatis, J. Bonca, and H. Q. Lin, Phys. Rev. Lett. \textbf{92}, 187601, (2004).
\bibitem{Czycholl} C. Schneider and G. Czycholl, Eur. Phys. J. B \textbf{64}, 43 (2008).
\bibitem{Snoke_1} D.W. Snoke, \textit{Coherence and Optical Emission from Bilayer Exciton Condensates}, Advances in Condensed Matter Physics \textbf{2011}, 938609 (2011). 
 \bibitem{Snoke_2} D. Snoke, Science \textbf{15}, 1368, (2002).
\bibitem{Tomio} Yuh Tomio, Kotaro Honda, and Tetsuo Ogawa, Phys. Rev. B \textbf{73}, 235108, (2006).
\bibitem{Micnas} R. Micnas, J. Ranninger, and S. Robaszkiewicz, Rev. Mod. Phys. \textbf{62}, 113 (1990).
\bibitem{Randeria} M. Randeria, in Bose-Einstein Condensation, edited by A. Griffin, D. W. Snoke, and S. Stringari (Cambridge University Press, Cambridge, U.K., 1995), p. 355.
\bibitem{Ohashi} Y. Ohashi and A. Griffin, Phys. Rev. Lett. \textbf{89}, 130402 (2002);
Phys. Rev. A \textbf{67}, 033603 (2003).
\bibitem{Falicov} L. M. Falicov and J. C. Kimball, Phys. Rev. Lett. \textbf{22}, 997 (1969).
\bibitem{Ramirez_1} R. Ramirez, L. M. Falicov, and J. C. Kimball, { Phys. Rev. B}, \textbf{2}, 3383 (1970).
\bibitem{Negele} J. W. Negele and H. Orland, \textit{Quantum Many-Particle Systems}, Addison-Wesley, Reading, MA, (1988).
\bibitem{Kopec_1} T. K. Kope\'{c}, Phys. Rev. B \textbf{73}, 132512 (2006).
\bibitem{Kopec_2} T. K. Kope\'{c}, Phys. Rev. B \textbf{73}, 104505 (2006).
\bibitem{Trombettoni} G. Grignani, A. Mattoni, P. Sodano, and A. Trombettoni, Phys. Rev. B \textbf{61}, 11676 (2000).
\bibitem{Abramovich} M. Abramovitz and I. Stegun, \textit{Handbook of Mathematical Functions} (Dover, New York, 1970).
\bibitem{Farkasovski_2} P. Farka\v{s}ovsk\'{y}, Phys. Rev. B \textbf{65}, 081102 (2002).
\bibitem{Popov} Popov, V. N., \textit{Functional integrals in quantum field theory and statistical physics} (D. Reidel Pub. Co., Dordrecht, Holland, 1983).
\bibitem{Thouless} D.J. Thouless, Ann. Phys. (N.Y.) \textbf{10}, 553, (1960).
\bibitem{Nozieres} P. Nozieres and Schmitt-Rink, J. Low Temp. Phys. \textbf{59}, 195 (1985).
\bibitem{Tsuchiya_1} Shunji Tsuchiya, Ryota Watanabe, and Yoji Ohashi, Phys. Rev. A \textbf{80}, 033613 (2009).
\bibitem{Tsuchiya_2} Ryota Watanabe, Shunji Tsuchiya, and Yoji Ohashi, Phys. Rev. A, \textbf{82}, 043630 (2010).
\bibitem{Kleinert} R.P. Feynman and H. Kleinert, Phys. Rev. A {\bf 34}, 5080 (1986).
\bibitem{Kim} S. Kim and M.Y. Choi, Phys. Rev. B {\bf 41}, 111 (1990).
\bibitem{Simanek} E. \v{S}im\'{a}nek, Phys. Rev. B {\bf 22}, 459 (1980).
\bibitem{Nakauchi} Y. Ohta, A. Nakauchi, R. Eder, K. Tsutsui, and S.
Maekawa, Phys. Rev. B \textbf{52}, 15617 (1995). 

\end{references}
\end{document}